\documentclass[a4paper, amsmath, mathtools, 11pt]{article}

\usepackage[dvipdfmx]{graphicx}

\usepackage{jcappub} 
\usepackage[top=30truemm,bottom=30truemm,left=20truemm,right=20truemm]{geometry}

\usepackage{eucal}
\usepackage{mathrsfs}
\usepackage{amsmath}

\usepackage[T1]{fontenc}
\usepackage[varg]{txfonts}

\allowdisplaybreaks[4]

\title{\boldmath Scalar field stochastic dynamics in de Sitter spacetime \\ from exact solutions of quantum deficient oscillators}

\author[\boldmath \spadesuit]{Yuta Nasuda,}
\author[\boldmath \clubsuit]{Koki Tokeshi,}
\author[\boldmath \spadesuit, \diamondsuit]{and Yuki Watanabe}

\affiliation[\boldmath \spadesuit]{\it School of Natural Sciences, National Institute of Technology (KOSEN), Gunma College, \\ 580 Toriba, Maebashi, Gunma 371-8530, Japan}
\vspace{0.2\baselineskip}
\affiliation[\boldmath \clubsuit]{\it Institute for Cosmic Ray Research (ICRR), The University of Tokyo, \\ 5-1-5 Kashiwanoha, Kashiwa, Chiba 277-8582, Japan}
\vspace{0.2\baselineskip}
\affiliation[\boldmath \diamondsuit]{\it Research Center for the Early Universe (RESCEU), The University of Tokyo, \\ 7-3-1 Hongo, Bunkyo, Tokyo 113-0033, Japan}

\emailAdd{y.nasuda.phys@gmail.com}
\emailAdd{tokeshi@icrr.u-tokyo.ac.jp}
\emailAdd{yuki.watanabe@gunma-ct.ac.jp}

\abstract{
	The stochastic dynamics of a scalar field in de Sitter spacetime can be regarded as a non-perturbative diffusion process, to which exact distribution and correlation functions are constructed by utilising the correspondence between diffusion and Schr\"{o}dinger equations. 
    The Krein--Adler transformation of the quantum harmonic oscillator deletes several pairs of the energy levels to define anharmonic oscillators that we dub quantum deficient oscillators, based on which this article constructs a new class of exact solutions in stochastic inflation. 
    In addition to the simplest single-well model, an exactly solvable double-well model is also presented. 
    The results are further extended to exactly solvable models with multiple wells, allowing analytical studies on various cosmological phenomenologies. 
}

\begin{document}

\maketitle
\flushbottom

\section{Introduction}
\label{sec:intro}

De Sitter spacetime has been extensively studied in theoretical physics. 
It describes the early-universe accelerated expanding phase dubbed \textit{cosmic inflation}~\cite{Starobinsky:1980te, Sato:1980yn, Guth:1980zm, Linde:1981mu, Albrecht:1982wi, Linde:1983gd} (see \textit{e.g.}~Refs.~\cite{Linde:2007fr, Baumann:2009ds, Sato:2015dga} for reviews), naturally providing the initial conditions required for the standard hot Big Bang scenario. 
Vacuum quantum fluctuations generated during inflation gave rise to all the present cosmological structures such as stars and galaxies~\cite{Mukhanov:1981xt}. 
Various large-scale observations such as Planck~\cite{Planck:2018jri, Planck:2018vyg, Planck:2019kim} have been confirming the nearly scale-invariant power spectrum of the adiabatic (curvature) primordial perturbation theoretically predicted by a simple ``vanilla'' inflationary model, where the vacuum energy of a single scalar field called the \textit{inflaton} dominates the total energy density of the universe. 

In addition to the inflaton field, various light scalar fields that do not contribute to the expansion of the universe are also motivated by high-energy physics, especially four-dimensional effective theories of inflation in string theory, see \textit{e.g.}~Ref.~\cite{Baumann:2014nda}. 
Those scalar fields are often called \textit{test} or \textit{spectator} fields, not only generating non-adiabatic (isocurvature) primordial fluctuations (and non-Gaussianities in some models) but also allowing us to probe de Sitter spacetime by neglecting the backreaction from the dynamics of the inflaton field to the spacetime geometry. 

If the fluctuations of the test fields are large, the non-perturbative analyses are necessary and can be achieved by the stochastic formalism of inflation~\cite{Starobinsky:1986fx}, \textit{stochastic inflation} for short, which describes the effective dynamics of the infrared large-scale fields beyond the standard perturbation theory (see Refs.~\cite{Starobinsky:1982ee, Sasaki:1987gy, Nambu:1987ef, Kandrup:1988sc, Nambu:1988je, Sasaki:1988df, Nakao:1988yi, Hosoya:1988yz, Nambu:1989uf, Morikawa:1989xz} for earlier works and Refs.~\cite{Mollerach:1990zf, Hattori:2005ac, Motohashi:2012bb, Assadullahi:2016gkk, Vennin:2016wnk, Pinol:2018euk, Pinol:2020cdp, Asadi:2023flu, Tada:2023fvd, Takahashi:2025hqt} for extensions and applications to multi-field models). 
The balance between the classical drift force for the large-scale fields and the stochastic inflow coming from the small-scale quantum fluctuations is non-perturbatively realised and gives rise to the stationary state~\cite{Starobinsky:1994bd}. 
It can be captured in stochastic inflation and is described by the stationary solution of the diffusion equation. 

Beyond the stationary limit, the stochastic formalism also enables us to keep track of a time-dependent evolution of a scalar field, governed by a stochastic differential equation, or equivalently by the Fokker--Planck equation~\cite{risken1989fpe} for the distribution function. 
While the number of exact solutions for the inflaton field is limited to the system with the quartic potential~\cite{Yi:1991ub} or the flat ultra-slow-roll potential~\cite{Pattison:2017mbe, Ando:2020fjm, Pattison:2021oen, Tada:2021zzj, Animali:2024jiz, Animali:2025pyf}, several exactly solvable systems can be constructed for a single test field, a subclass of which was very recently studied and summarised in Ref.~\cite{Honda:2024evc}. 

The availability of those exact solutions can be systematically understood from the viewpoint of the correspondence between the diffusion equation in stochastic inflation and the Schr\"{o}dinger equation in non-relativistic quantum mechanics. 
In addition to the quantum harmonic oscillator solved in terms of the Hermite polynomials, there are other systems described in terms of the other classical orthogonal polynomials, such as the (associated) Laguerre polynomials and Jacobi polynomials. 
Those systems are solved in a consistent manner, called the factorisation method~\cite{RevModPhys.23.21}.
They are endowed with an underlying symmetry, often referred to as \textit{shape invariance}~\cite{Gendenshtein:1983skv}, found in the context of supersymmetric quantum mechanics~\cite{Witten:1981nf}.
This symmetry gives a sufficient condition for a quantum-mechanical system to be exactly solvable.
Ten quantum-mechanical potentials have been found in total up to today that respect shape invariance and are solved in terms of the classical orthogonal polynomials, following the classification presented in Ref.~\cite{Cooper:1994eh}. 
There are, however, other exactly solvable ones called Natanzon potentials~\cite{Natanzon:1979sr, Ginocchio:1984ih, PhysRevD.36.2458}, which are exceptional in the sense that they do not respect shape invariance. The Schr\"{o}dinger equations with those potentials can be solved exactly in terms of hypergeometric functions. 

Interestingly, a no-go statement presently known as Bochner's theorem~\cite{1884_Routh, bochner1929} tells us that there is no essentially new class of exactly solvable potentials in quantum mechanics in terms of the classical orthogonal polynomials.
In other words, Bochner's theorem gives all classical exact solutions to second-order differential equations with polynomial eigenfunctions that correspond to the conventional shape-invariant potentials.
The assumptions in the theorem must therefore be loosened in order to go beyond the classical exact solutions. 
This led to the construction of the ($q$-)Askey--Wilson scheme~\cite{Andrews_Askey_Roy_1999, Ismail_2005, KoekoekpFq2010}, which is an organisation of the (basic or $q$-)hypergeometric polynomials into a hierarchy. 
It opened a new window for finding a new class of orthogonal polynomials including discrete quantum mechanics~\cite{Odake:2007cs, Odake:2008bf, Odake:2011gk}. 
Several authors also considered piecewise analytic potentials (for recent works, see \textit{e.g.}~Refs.~\cite{Sasaki:2016ekz, Znojil:2016elp, Sasaki:2022aby, Nasuda:2023ahq,doi:10.1142/S0219887825400304}), in which the wavefunction can be obtained analytically but is defined domain by domain.

Moreover, a class of novel exact wavefunctions can still be constructed in the framework of one dimensional continuous quantum mechanics keeping the potential, wavefunctions, and their derivatives continuous by the so-called Darboux transformation~\cite{darboux}. 
It can be regarded as a generalisation of Crum's transformation~\cite{10.1093/qmath/6.1.121} used to construct supersymmetric partners for a given potential. 
One leading example of the transformation is the Krein--Adler transformation~\cite{ctx59091312420007871, Adler1994}.
A well-known exactly solvable potential such as the quantum harmonic oscillator is transformed in such a way that the newly constructed potential possesses a set of \textit{deficient} energy levels, generating a new set of orthogonal polynomials. 
Most of the energy eigenvalues in the original system typically remain unchanged, while several pairs of neighbouring eigenvalues are deleted in the transformed system (see later Figure~\ref{fig:rem}). 
Such models considered in this article are therefore dubbed \textit{quantum deficient oscillators}. 

This work utilises some of those potentials and explicitly constructs the exact wavefunctions in an elementary way, for which the Krein--Adler scheme provides a systematic and unified understanding. 
By mapping those exact solutions in quantum mechanics to the diffusion system describing the scalar-field dynamics in stochastic inflation, we construct novel exact solutions for the scalar field in de Sitter spacetime, which keep track of whole the time evolution of the scalar field. 
Those also enable us to derive the time-dependent correlation functions, \textit{i.e.}~statistical moments at arbitrary order. 

The present article is organised as follows. 
Section~\ref{sec:sinf} reviews the dynamics of a scalar field in de Sitter spacetime described in the framework of stochastic formalism of inflation, the mapping between the Fokker--Planck equation and the Schr\"{o}dinger equation, and the simplest exactly solvable model such that the scalar field is confined in the quadratic potential. 
Section~\ref{sec:esnv} then introduces a few exactly solvable models in non-relativistic quantum mechanics, in which the potentials are deformed from the harmonic oscillator and do not respect shape invariance. 
After the two anharmonic potentials are introduced and the corresponding exact wavefunctions are constructed by an elementary series-solution method, the relation to and construction based on a scheme of the Krein--Adler transformation are briefly summarised. 
Section~\ref{sec:statds} maps them back on the stochastic-inflation side and provides the novel and fully time-dependent exact solutions that describe the scalar-field dynamics in de Sitter spacetime. 
Finally, Section~\ref{sec:disc} is devoted to summarise the present work and to mention its possible applications and extensions. 
The natural system of units is used throughout and the reduced Planck mass, $M_{\rm P} \simeq 2.4 \times 10^{18} \, \mathrm{GeV}$, is set to unity unless otherwise stated. 

\section{Scalar field dynamics in de Sitter spacetime}
\label{sec:sinf}

The existence of a nearly constant horizon characterises de Sitter spacetime. 
Small-scale quantum fluctuations are stretched due to the expansion of the universe to cross out the horizon, losing their quantumness and becoming a part of large-scale classical fluctuations. 
From the viewpoint of the large-scale fluctuations, their dynamics is affected by the continuous but random inflow due to the small-to-large-scale transition. 
The resultant effective dynamics of the large-scale modes then becomes stochastic, and the stochastic formalism of inflation precisely describes this random motion of the large-scale field configurations~\cite{Starobinsky:1986fx}. 

This section aims at providing the basic equations in stochastic inflation in the former half, putting in Section~\ref{subsec:dfsh} a special emphasis on the correspondence between the diffusion system and non-relativistic quantum mechanics. 
Section~\ref{subsec:hodemo} is based on this correspondence and demonstrates a construction of the simplest exact solution in stochastic inflation, namely a test field in the quadratic potential in the expanding universe, by which readers get ready to construct more complicated exact solutions in stochastic inflation in the remaining sections. 

\subsection{Stochastic inflation for test field}

With a flat Friedmann--Lema\^{i}tre--Robertson--Walker (FLRW) metric, $\dd s^{2} = - \dd t^{2} + [ a (t) ]^{2} \, \dd \vb*{x}^{2}$, the relevant terms in the Lagrangian that the present article focusses on are 
\begin{equation}
	\mathcal{L} 
	\supset \frac{R}{2} - \frac{1}{2} g^{\mu \nu} ( \partial_{\mu} \phi ) ( \partial_{\nu} \phi ) - V (\phi) 
	\,\, . 
\end{equation}
The real scalar field $\phi$ is assumed to be different from the inflaton field that drives inflation, and it is rather regarded as a test field in de Sitter spacetime throughout. 
The Klein--Gordon field equation for the scalar field is given by $\ddot{\phi} + 3 H \dot{\phi} - \nabla^{2} \phi / a^{2} + \dd V / \dd \phi = 0$, where $a = a (t)$ is the scale factor, $H \coloneqq \dd \ln a (t) / \dd t$ is the Hubble parameter assumed to be constant for considering the dynamics of $\phi$, and $V = V (\phi)$ is the potential on which $\phi$ rolls. 
It is more convenient for our purpose to rewrite this in terms of the number of $e$-folds defined through $\dd N = H \, \dd t$ as  
\begin{equation}
	\pdv[2]{\phi}{N} + 3 \pdv{\phi}{N} - \frac{\nabla^{2} \phi}{(a H)^{2}} + \frac{1}{H^{2}} \dv{V}{\phi} = 0 
	\,\, . 
\end{equation}

From now on, let us review the stochastic formalism of inflation---an effective and non-perturbative field-theoretical approach to large-scale fluctuations---from its heuristic but practical construction that allows easier physical interpretation, instead of deriving the same effective equation from a more robust influence functional method~\cite{Morikawa:1989xz, Hu:1994dka, Matarrese:2003ye, PerreaultLevasseur:2013kfq, Moss:2016uix, Tokuda:2017fdh, Tokuda:2018eqs}. 
The central idea to construct the formalism is to decompose the whole field into the long-wavelength (super-horizon, infrared) and the short-wavelength (sub-horizon, ultraviolet) modes as 
\begin{equation}
	\phi (N, \, \vb*{x}) 
	= \int \frac{\dd^{3} k}{(2 \pi)^{3}} \, \widetilde{\phi} (N, \, \vb*{x}) e^{i \vb*{k} \cdot \vb*{x}} 
	= \phi_{\rm L} (N, \, \vb*{x}) + \phi_{\rm S} (N, \, \vb*{x}) 
	\,\, . 
\end{equation}
The ``L''-field (``S''-field) accumulates all the $k$-modes that are longer (shorter) than the rescaled horizon scale, $k_{\sigma} (N) \coloneqq \sigma a(N) H$ with $0 < \sigma \ll 1$. 
The infrared modes are assumed to be well classicalised outside the horizon, while the ultraviolet modes describe the quantum fluctuations. 
Each component is therefore defined in Fourier space by 
\begin{subequations}
	\label{eq:pre_dcp_phi}
	\begin{align}
		\phi_{\rm L} (N, \, \vb*{x}) 
		&\coloneqq \int \frac{\dd^{3} k}{(2 \pi)^{3}} \,  \widetilde{\phi} (N, \, \vb*{k}) \Theta [ - k + k_{\sigma} (N) ] e^{i \vb*{k} \cdot \vb*{x}} \,\, , 
		\label{eq:pre_dcp_phi1}
		\\ 
		\phi_{\rm S} (N, \, \vb*{x}) 
		&\coloneqq \int \frac{\dd^{3} k}{(2 \pi)^{3}} \,  \widetilde{\phi} (N, \, \vb*{k}) \Theta [ + k - k_{\sigma} (N) ] e^{i \vb*{k} \cdot \vb*{x}} \,\, . 
		\label{eq:pre_dcp_phi2}
	\end{align}
\end{subequations}
The cut-off window function can in principle be chosen arbitrarily as long as it goes to unity for a small argument and vanishes for a large argument, but, for simplicity, the Heaviside step function is implemented throughout this article following the literature. 
This specific choice of the window function allows the following discussion to be simplified; see Refs.~\cite{Hu:1992ig, Casini:1998wr, Winitzki:1999ve, Matarrese:2003ye, Liguori:2004fa, Breuer:2006cd, Mahbub:2022osb} for smooth filtering and coloured noise. 

It is more convenient to express the field equation in Hamiltonian formulation as
\begin{subequations}
	\label{eq:pre_ham}
	\begin{align}
		\dv{\phi}{N} 
		&= \pi 
		\,\, , \\ 
		0 
		&= \dv{\pi}{N} + 3 \pi - \frac{\nabla^{2} \phi}{(a H)^{2}} + \frac{1}{H^{2}} \dv{V}{\phi} 
		\,\, . 
	\end{align}
\end{subequations}
The conjugate momentum $\pi$ is also decomposed into its large-scale and small-scale modes in the same way as Eqs.~(\ref{eq:pre_dcp_phi}). 
Substitution of the L/S decompositions into Eqs.~(\ref{eq:pre_ham}) yields 
\begin{subequations}
	\label{eq:pre_ham_dcp}
	\begin{align}
		\dv{\phi_{\rm L}}{N} - \pi_{\rm L} 
		&= - \dv{\phi_{\rm S}}{N} + \pi_{\rm S} 
		\,\, ,\\ 
		\dv{\pi_{\rm L}}{N} + 3 \pi_{\rm L} + \frac{1}{H^{2}} \dv{V}{\phi}{} (\phi_{\rm L}) 
		&= - \dv{\pi_{\rm S}}{N} - 3 \pi_{\rm S} + \frac{\nabla^{2} \phi_{\rm S}}{(a H)^{2}} - \frac{1}{H^{2}} \dv[2]{V}{\phi} {} (\phi_{\rm L}) \cdot \phi_{\rm S}
		\,\, .  
	\end{align}
\end{subequations}
Note that in the second equation the spatial gradient for the long-wavelength mode has been neglected, and the potential is expanded around $\phi = \phi_{\rm L}$, \textit{i.e.}~$V' (\phi) = V' (\phi_{\rm L}) + V'' (\phi_{\rm L}) \cdot \phi_{\rm S} + \mathcal{O} (\phi_{\rm S}^{2})$ (recall that $\phi_{\rm L} + \phi_{\rm S} = \phi$). 
Since the effective equation of motion for the long-wavelength mode is desired here, let us substitute Eq.~(\ref{eq:pre_dcp_phi2}) into Eqs.~(\ref{eq:pre_ham_dcp}) to get 
\begin{subequations}
	\begin{align}
		\dv{\phi_{\rm L}}{N} - \pi_{\rm L} 
		&= - \int \frac{\dd^{3} k}{(2 \pi)^{3}} \, \pdv{N} \qty{ 
		  \widetilde{\phi} (N, \, \vb*{k}) \Theta [ + k - k_{\sigma} (N) ] 
        } e^{i \vb*{k} \cdot \vb*{x}} 
        + \int \frac{\dd^{3} k}{(2 \pi)^{3}} \,  \widetilde{\pi} (N, \, \vb*{k}) \Theta [ + k - k_{\sigma} (N) ] e^{i \vb*{k} \cdot \vb*{x}}
		\notag \\ 
		&=  \int \frac{\dd^{3} k}{(2 \pi)^{3}} \, \widetilde{\phi} (N, \, \vb*{k}) 
		\cdot \dv{k_{\sigma} (N)}{N} \cdot 
		\delta_{\rm D} [ + k - k_{\sigma} (N) ] e^{i \vb*{k} \cdot \vb*{x}} 
		\eqqcolon n_{\phi} (N, \, \vb*{x}) 
		\,\, \label{eq:pre_ham_dcp_a}\\ 
		\dv{\pi_{\rm L}}{N} + 3 \pi_{\rm L} + \frac{1}{H^{2}} \dv{V}{\phi} {} (\phi_{\rm L}) 
		&= - \dv{\pi_{\rm S}}{N} - 3 \pi_{\rm S} + \frac{\nabla^{2} \phi_{\rm S}}{(a H)^{2}} - \frac{1}{H^{2}} \dv[2]{V}{\phi} {} (\phi_{\rm L}) \cdot \phi_{\rm S}
		\notag \\ 
		&= \int \frac{\dd^{3} k}{(2 \pi)^{3}} \, \widetilde{\pi} (N, \, \vb*{k}) 
		\cdot \dv{k_{\sigma} (N)}{N} \cdot 
		\delta_{\rm D} [ + k - k_{\sigma} (N) ] e^{i \vb*{k} \cdot \vb*{x}} 
		\eqqcolon n_{\pi} (N, \, \vb*{x}) 
		\,\, . \label{eq:pre_ham_dcp_b}
	\end{align}
\end{subequations}
Due to $\widetilde{\pi} (N, \, \vb*{k}) = \partial \widetilde{\phi} (N, \, \vb*{k}) / \partial N$, the terms in Eq.~(\ref{eq:pre_ham_dcp_a}) including the Heaviside function have cancelled each other out except for the contribution of the Dirac delta function. 
Most of the terms in the right-hand side of Eq.~(\ref{eq:pre_ham_dcp_b}) have vanished because of the linearised equation of motion~(\ref{eq:linear}) that we see later.
The noises $n_{\phi}$ and $n_{\pi}$ of the scalar field and its conjugate momentum, respectively, have been introduced. 
The reason why those are called noises will become clear soon. 
It can be seen at this point that not only the classical potential drift but also the noise, where only the mode with $k = k_{\sigma} (N)$ at each time exists, altogether affect the effective dynamics of the long-wavelength mode. 

The sub-horizon, hence quantum, fluctuations $\phi_{\rm S}$ and $\pi_{\rm S}$ are expressed in terms of their creation and annihilation operators as 
\begin{subequations}
	\label{eq:pre_modep}
	\begin{align}
		\phi_{\rm S} (N, \, \vb*{x}) 
		&= \int \frac{\dd^{3} k}{(2 \pi)^{3}} \, 
		\Theta [ + k - k_{\sigma} (N) ] 
		[ 
			a (\vb*{k}) \widetilde{\phi}_{k} (N) e^{+ i \vb*{k} \cdot \vb*{x}} + a^{\dagger} (\vb*{k}) \widetilde{\phi}^{\star}_{k} (N) e^{- i \vb*{k} \cdot \vb*{x}} 
		] 
		\,\, , 
		\\ 
		\pi_{\rm S} (N, \, \vb*{x}) 
		&= \int \frac{\dd^{3} k}{(2 \pi)^{3}} \, 
		\Theta [ + k - k_{\sigma} (N) ] 
		[ 
			a (\vb*{k}) \widetilde{\pi}_{k} (N) e^{+ i \vb*{k} \cdot \vb*{x}} + a^{\dagger} (\vb*{k}) \widetilde{\pi}^{\star}_{k} (N) e^{- i \vb*{k} \cdot \vb*{x}} 
		] 
		\,\, . 
	\end{align}
\end{subequations}
In Eqs.~(\ref{eq:pre_modep}), stars and daggers denote the complex and Hermitian conjugates, respectively. 
Since the stochastic formalism treats the large-scale classicalised fluctuations non-perturbatively while the standard linear perturbation theory applies to the small-scale quantum fluctuations, the mode function $\widetilde{\phi}_{k} (N)$ satisfies the linearised equation of motion
\begin{equation}\label{eq:linear}
	\dv[2]{\widetilde{\phi}_{k}}{N} 
	+ 3 \dv{\widetilde{\phi}_{k}}{N} 
	+ \qty[ 
		\qty( \frac{k}{a H} )^{2} + \frac{1}{H^{2}} \dv[2]{V}{\phi} {} (\phi_{\rm L}) 
	] \, \widetilde{\phi}_{k} 
	= 0 
	\,\, . 
\end{equation}
This can be reduced to the Hankel differential equation, as usual, of order $\nu \coloneqq (3/2) \sqrt{ 1 - 4 V'' (\phi_{\rm L}) / 9 H^{2} }$, and the mode function is then given by 
\begin{equation}
	\widetilde{\phi}_{k} (N) 
	= (- k \tau)^{3/2} [ 
		c_{1} (k) H_{\nu}^{(1)} (- k \tau) + c_{2} (k) H_{\nu}^{(2)} (- k \tau) 
	] 
	\,\, . 
	\label{eq:pre_mdsol}
\end{equation}
In Eq.~(\ref{eq:pre_mdsol}), $H_{\nu}^{(1)} (z)$ and $H_{\nu}^{(2)} (z)$ are the Hankel functions of the first and second kinds of order $\nu$, respectively. 
The Bunch--Davies vacuum initial condition sets the coefficients $c_{i} (k)$ and the mode function is then determined. 

The vacuum expectation value of the noises can now be calculated and expressed in terms of the mode functions as
\begin{align}
    \langle 
		\vb*{n} (N_{1}, \, \vb*{x}_{1}) \otimes \vb*{n} (N_{2}, \, \vb*{x}_{2}) 
	\rangle 
	&= \frac{k^{3}}{2 \pi^{2}} \mqty( 
		\displaystyle | \widetilde{\phi}_{k} |^{2} 
		& \displaystyle \widetilde{\phi}_{k} \pdv{ \widetilde{\phi}_{k}^{\star} }{N} 
		\\[1.5ex] 
		\displaystyle \pdv{ \widetilde{\phi}_{k} }{N} \widetilde{\phi}_{k}^{\star}  
		&\displaystyle \abs{ \pdv{ \widetilde{\phi}_{k} }{N} }^{2} 
	) 
	\, 
	\frac{\sin (k \abs{ \vb*{x}_{1} - \vb*{x}_{2} })}{ k \abs{ \vb*{x}_{1} - \vb*{x}_{2} } } 
	\, 
	\delta_{\rm D} (N_{1} - N_{2}) 
	\eval{}_{k = k_{\sigma} (N)} 
	\notag \\ 
	&= \qty( \frac{H}{2 \pi} )^{2} \frac{4 [ \Gamma (\nu) ]^{2}}{\pi} \mqty( 
		1 & \nu - 3/2 \\ \nu - 3/2 & (\nu - 3/2)^{2} 
	) 
	\qty( \frac{\sigma}{2} )^{3 - 2 \nu} 
	\, 
	\frac{\sin (k \abs{ \vb*{x}_{1} - \vb*{x}_{2} })}{ k \abs{ \vb*{x}_{1} - \vb*{x}_{2} } } 
	\, 
	\delta_{\rm D} (N_{1} - N_{2}) 
	\,\, , 
\end{align}
where $\vb*{n} \coloneqq (n_{\phi}, \, n_{\pi})^{\mathsf{T}}$ and $\Gamma(\nu)$ is the Gamma function. 
In the $\nu \to 3/2$ (massless) limit, only the noise of $\phi$ survives, while those involving $\pi$ vanish.  
This completely characterises the statistical properties of the noise. 
Therefore, the effective equation of motion in the slow-roll regime is given by (hereafter $\phi_{\rm L}$ is denoted simply by $\phi$) 
\begin{equation}
	\dv{\phi}{N}
	= - \frac{1}{3 H^{2}} \dv{V}{\phi} + \frac{H}{2 \pi} \xi (N) 
	\,\, . 
	\label{eq:pre_lan}
\end{equation}
Here, the random and normalised Gaussian noise has been denoted as $\xi = \xi (N)$. 
It has the following correlation properties: 
\begin{subequations}
	\begin{align}
		\expval{ \xi (N, \, \vb*{x}) } 
		&= 0 
		\,\, , 
		\\ 
		\expval{ \xi (N_{1}, \, \vb*{x}_{1}) \xi (N_{2}, \, \vb*{x}_{2}) } 
		&= \frac{ \sin ( k_{\sigma} \abs{ \vb*{x}_{1} - \vb*{x}_{2} } ) }{ k_{\sigma} \abs{ \vb*{x}_{1} - \vb*{x}_{2} } } 
		\delta_{\rm D} (N_{1} - N_{2}) 
		\xrightarrow{\vb*{x}_{1} \to \vb*{x}_{2}} 
		\delta_{\rm D} (N_{1} - N_{2}) 
		\,\, . 
	\end{align}
\end{subequations}
The noise originates from the sub-horizon fluctuations. Their initial condition is given by the standard Bunch--Davies vacuum. 
In what follows, the one-point statistics is of interest, so the limit $\vb*{x}_{1} \to \vb*{x}_{2}$ is relevant, which gives rise to $\expval{ \xi (N_{1}) \xi (N_{2}) } = \delta_{\rm D} (N_{1} - N_{2})$, \textit{i.e.}~a time and another time are uncorrelated. 
The Langevin equation (\ref{eq:pre_lan}) describes the evolution of each coarse-grained Hubble region in this limit. 

\subsection{Fokker--Planck and Schr\"{o}dinger equations}
\label{subsec:dfsh}

The fact that the scalar field $\phi$ evolves stochastically subjected to the uncoloured noise $\xi$ allows us to derive the evolution equation, the Fokker--Planck equation~\cite{risken1989fpe}, for the probability distribution function (PDF) of $\phi$ at a given time $N$, 
\begin{equation}
	\pdv{f}{N} 
	= \frac{1}{3 H^{2}} \pdv{\phi} \qty( \dv{V}{\phi} f ) + \frac{H^{2}}{8 \pi^{2}} \pdv[2]{f}{\phi} 
	\eqqcolon \mathcal{L}_{\rm FP} f 
	\,\, , 
	\qquad 
	f = f (\phi, \, N) = f (\phi, \, N \mid \phi_{0}, \, N_{0}) 
	\,\, , 
	\label{eq:pre_fp}
\end{equation}
where $\phi_0$ and $N_0$ are initial field value and time, respectively.
An immediate consequence is that the stationary solution to Eq.~(\ref{eq:pre_fp}), satisfying $\partial f_{\infty} (\phi) / \partial N = 0$ for which the probability current vanishes, is given by~\cite{Starobinsky:1994bd}
\begin{equation}
	f_{\infty} (\phi) 
	\coloneqq \lim_{N \to \infty} f (\phi, \, N) 
	= 
	\frac{
		\displaystyle \exp \qty[ - \frac{8 \pi^{2}}{3 H^{4}} V (\phi) ] 
	}{ 
		\displaystyle \int \dd \phi \, \exp \qty[ - \frac{8 \pi^{2}}{3 H^{4}} V (\phi) ] 
	} 
	\,\, 
	\label{eq:pre_dist_eqb}
\end{equation}
for arbitrary potentials $V(\phi)$.
The stationary one-point correlation function is associated with the stationary distribution, and is given by 
\begin{equation}
	\expval{ \phi^{n} }_{\infty} 
	\coloneqq \lim_{N \to \infty} \expval{ \phi^{n} } (N) 
	= \int \dd \phi \, \phi^{n} f_{\infty} (\phi) 
	\,\, . 
\end{equation}

The stationary state is a result of the balance between the classical drift and the quantum diffusion, the non-perturbative dynamics captured in the stochastic formalism, which therefore cannot be apprehended in the standard perturbation theory, encountering the secular divergence of the statistical quantities~\cite{Starobinsky:1994bd, Tsamis:2005hd, Kahya:2006hc, Prokopec:2007ak, Finelli:2008zg, Tokuda:2017fdh, Tokuda:2018eqs, Gorbenko:2019rza, Kamenshchik:2021tjh} (see also Refs.~\cite{Kamenshchik:2024ybm, Kamenshchik:2025ses} for alternative approaches) unless one relies on some resummation scheme~\cite{Honda:2023unh}. 
When one is interested in keeping track of the time evolution of the scalar field, on the other hand, it is often challenging to solve a partial differential equation such as the Fokker--Planck equation even numerically, in particular in multi-field models. 

However, in the present article, the dynamics of a single field is of interest and there are a few well-known exact solutions by which the time evolution of the distribution function of the stochastic field $\phi$ can be kept track of analytically. 
The Fokker--Planck equation can be solved exactly when the potential is, for instance, purely quadratic, \textit{i.e.}~$V (\phi) = m^{2} \phi^{2} / 2$. 
The corresponding stochastic dynamics is known as the Ornstein--Uhlenbeck process. 
Other potentials for which the Fokker--Planck equation can be solved exactly can be found systematically by mapping it to the Schr\"{o}dinger equation. A class of them is summarised in Ref.~\cite{Honda:2024evc}. 
The construction procedure starts with the method of spectral decomposition of $f$ (see \textit{e.g.}~Refs.~\cite{Starobinsky:1986fx, Starobinsky:1994bd, Markkanen:2019kpv, Panagopoulos:2019ail, Giudice:2021viw, Achucarro:2021pdh, Honda:2024evc, Tokeshi:2024kuv} for the use of the same technique in the stochastic-inflation context), given by
\begin{equation}
	f (\phi, \, N) 
	= \sum_{n = 0}^{\infty} c_{n} \exp \qty[ - \frac{4 \pi^{2}}{3 H^{4}} V (\phi) ] \Psi_{n} (\phi) T_{n} (N) 
	\,\, . 
    \label{eq:pre_sdcmp}
\end{equation}
Here, the non-existence of continuous part is assumed. 
In other words, for the domain $\phi \in (\phi_{1}, \, \phi_{2})$ in which the potential is defined, it is assumed that the field $\phi$ is bounded from both sides, that is, $V (\phi) \to \infty$ as $\phi \to \phi_{1}$ \textit{and} $\phi \to \phi_{2}$.

The coefficient $c_{n}$ is determined by an initial distribution at $N = N_{0}$, which, throughout this article, is assumed to be the Dirac $\delta$-function: 
\begin{equation}\label{eq:initial_dist}
f (\phi, \, N = N_{0}) = \delta_{\rm D} (\phi - \phi_{0}) \, .
\end{equation}
The time component $T_{n} (N)$ is immediately determined as $T_{n} = e^{- \lambda_{n} (N - N_{0})}$ up to a proportional constant, which is absorbed by $c_{n}$. 
The function $\Psi_{n} (\phi)$ turns out to be the wavefunction governed by the stationary Schr\"{o}dinger equation
\begin{equation}
	\qty[ - \frac{H^{2}}{8 \pi^{2}} \dv[2]{\phi} + V_{\rm S} (\phi) ] \Psi_{n} (\phi) 
	= \lambda_{n} \Psi_{n} (\phi) 
	\,\, , 
	\qquad 
	V_{\rm S} (\phi) 
	\coloneqq \frac{2 \pi^{2}}{9 H^{6}} \qty( \dv{V}{\phi} )^{2} - \frac{1}{6 H^{2}} \dv[2]{V}{\phi} 
	\,\, . 
	\label{eq:pre_spexps}
\end{equation}
It becomes $( - \dd^{2} / \dd y^{2} + V_{\rm S} ) \Psi_{n} = \lambda_{n} \Psi_{n}$ for the non-dimensional field variable defined by $y \coloneqq \sqrt{8 \pi^{2} / H^{2}} \, \phi$.  
For an exactly solvable potential in non-relativistic quantum mechanics, the corresponding analytical distribution function as solutions to the Fokker--Planck equation can therefore be constructed. 

A direct approach is to start from an exactly solvable quantum-mechanical potential $V_{\rm S} (\phi)$ in Eq.~(\ref{eq:pre_spexps}) and then to the Riccati equation to get the corresponding exactly solvable potential $V (\phi)$ for the scalar field in de Sitter spacetime. 
However, in general, this procedure comes with an integration constant that must be chosen appropriately to avoid the singular behaviour of $V (\phi)$. 
In order not to face with this subtlety, the potential $W (\phi)$ will be introduced later, which is uniquely determined through the logarithmic derivative of the ground-state solution to Eq.~(\ref{eq:pre_spexps}). 
As will be seen in Section~\ref{sec:statds}, the potential $V (\phi)$ is given simply in terms of the ground-state wavefunction accordingly. 

\subsection{Quantum harmonic oscillator: exact distribution function}
\label{subsec:hodemo}

The most famous situation in which the system can be solved exactly is the scalar field in the quadratic potential. 
Through this example, let us demonstrate the entire procedure to construct an exact solution in stochastic inflation, \textit{i.e.}~the exact formulas for both the distribution and correlation functions (statistical moments). 
Let us consider the Schr\"{o}dinger potential\footnote{
    The replacement $A^{1/2} \to \omega / 2$ reproduces the expressions in Ref.~\cite{Honda:2024evc}. 
}
\begin{equation}
	V_{\rm S} (\phi) 
	= A y^{2} - \sqrt{A} 
	= \sqrt{A} \, (z^{2} - 1) 
        \eqqcolon V_{\rm HO}(\phi)
	\,\, , 
	\qquad 
	z = A^{1/4} y 
	\,\, , 
    \qquad 
    A > 0 
    \,\, . 
	\label{eq:pre_ex_hopot}
\end{equation}
The Schr\"{o}dinger equation in terms of $z$ then reads $[ - \dd^{2} / \dd z^{2} + (z^{2} - 1) ] \Psi_{n} (\phi) = (\lambda_{n} / \sqrt{A}) \Psi_{n} (\phi)$, with the eigenenergy being 
\begin{equation}
    \frac{ \lambda_{n} }{ \sqrt{A} } 
    = 2 n 
    \,\, , 
    \qquad 
    n = 0, \, 1, \, \dots 
    \,\, . 
    \label{eq:pre_ex_hoev}
\end{equation}

The wavefunction is then given in terms of the Hermite polynomial, such that $\Psi_{n} (\phi) = e^{- z^{2} / 2} H_{n} (z)$. 
The orthogonal relation among the Hermite polynomials, 
\begin{equation}
    \int_{\mathbb{R}} \dd z \, e^{- z^{2}} H_{n} (z) H_{m} (z) 
    = 2^{n} n! \sqrt{\pi} \, \delta_{nm} \,\, , 
    \label{eq:pre_ex_ortho}
\end{equation}
where $\delta_{nm}$ is the Kronecker delta, leads to the properly normalised wavefunction
\begin{equation}
	\widehat{\Psi}_{n} (\phi) 
	= \frac{A^{1/8}}{\sqrt{2^{n} n! \sqrt{\pi}}} \qty( \frac{8 \pi^{2}}{H^{2}} )^{1/4} \exp \qty( - \frac{z^{2}}{2} ) H_{n} (z) 
	\,\, . 
    \label{eq:pre_dem_nomwf}
\end{equation}
Hereafter, a hat $\widehat{ \bullet }$ indicates that the quantity $\bullet$ is properly normalised according to the normalisation condition $\displaystyle \int \dd \phi \, | \widehat{ \Psi }_{n} (\phi) |^{2} = 1$. 

From the normalised ground-state wavefunction, the potential of the scalar field in de Sitter spacetime is obtained as
\begin{equation}
	\frac{V (\phi)}{H^{4}} 
	= \frac{3}{4 \pi^{2}} \cdot \frac{z^{2}}{2} 
	\,\, . 
    \label{eq:pre_ex_infv}
\end{equation}
See Eq.~(\ref{eq:cat_spotinf}) in Section~\ref{sec:statds} for the general formula. 
The potential for the scalar field in de Sitter spacetime is, therefore, also quadratic in this case, which explains why everything is Gaussian to be solved exactly. 

The orthogonality relation (\ref{eq:pre_ex_ortho}) also enables us to determine the coefficient $c_{n}$ under the initial distribution~(\ref{eq:initial_dist}), and together with the temporal function $T_{n} (N)$ the time-dependent distribution function reads 
\begin{subequations}
\begin{align}
	f (\phi, \, N) 
	&= \frac{A^{1/4}}{\sqrt{\pi}} \sqrt{ \frac{8 \pi^{2}}{H^{2}} } \,  
	\exp \qty( - z^{2} ) 
	\sum_{n = 0}^{\infty} \frac{1}{2^{n} n!} H_{n} (z) H_{n} (z_{0}) e^{- 2 n \sqrt{A} \, (N - N_{0})} 
    \label{eq:pre_dem_distf}
    \\ 
	&= \frac{1}{\sqrt{ 2 \pi \sigma^{2} (N) }} \, \exp \qty{ 
		- \frac{ \qty[ \phi - \mu (N) ]^{2} }{ 2 \sigma^{2} (N) } 
	} 
	\,\, , 
\end{align}
\end{subequations}
where the Mehler summation formula~\cite{Andrews_Askey_Roy_1999} has been used in going to the second line and is given by 
\begin{equation}
	\sum_{n = 0}^{\infty} 
	\frac{\alpha^{n}}{n!} H_{n} (z_{1}) H_{n} (z_{2}) 
	= \frac{1}{\sqrt{1 - 4 \alpha^{2}}} \, \exp \qty{ 
		\frac{4 \alpha}{1 - 4 \alpha^{2}} \qty[ 
			z_{1} z_{2} - \alpha (z_{1}^{2} + z_{2}^{2}) 
		] 
	} 
	\,\, . 
    \label{eq:pre_ex_mehler}
\end{equation}
The distribution is a Gaussian function, where the mean $\mu \coloneqq \langle \phi \rangle$ and variance $\sigma^{2} \coloneqq \langle \phi^{2} \rangle - \langle \phi \rangle^{2}$ are given by 
\begin{subequations}
	\begin{align}
		\mu (N) 
		&= \phi_{0} 
		\exp [ - 2 \sqrt{A} \, (N - N_{0}) ] 
		\,\, , 
		\\ 
		\sigma^{2} (N) 
		&= \frac{1}{4 \sqrt{A}} \qty( \frac{H}{2 \pi} )^{2} \qty{ 
			1 - \exp [ - 4 \sqrt{A} \, (N - N_{0}) ] 
		} 
		\,\, . 
	\end{align}
\end{subequations}
In this simplest case, the infinite summation in the spectral decomposition has been reduced to a closed form, which is confirmed to be a Gaussian function as expected. 
However, closed-form expressions cannot be found in general. 
In such cases, the exact distribution and correlation functions may be approximated by truncating the series at a finite order, which sometimes gives a good approximation due to the exponentially decaying factor that becomes negligible as time goes by. 

The wavefunctions and distribution function in this quantum harmonic oscillator case will be referred to as our benchmark in Section~\ref{sec:esnv}. 

\section{Exactly solvable Krein--Adler-transformed anharmonic oscillator}
\label{sec:esnv}

Having mapped the Fokker--Planck equation to the Schr\"{o}dinger equation and demonstrated the simplest case, it is now in a good position to introduce several more complicated potentials that admit exact solutions in quantum mechanics. 
The most famous exact solution in non-relativistic quantum mechanics is the harmonic oscillator, in which the wavefunctions are given in terms of the Hermite polynomials as written in elementary textbooks and reviewed in Section~\ref{subsec:hodemo}. 
There are other exactly solvable quantum-mechanical systems solved by the other two classical orthogonal polynomials, namely the associated Laguerre polynomials (related to \textit{e.g.}~hydrogen atom) and the Jacobi polynomials (that describe a system with \textit{e.g.}~the Rosen--Morse potential). 

The quest for exact solutions in quantum mechanics has a relatively long history, and all the exactly solvable potentials in terms of the classical orthogonal polynomials known today are systematically classified in \textit{e.g.}~Refs.~\cite{RevModPhys.23.21, Cooper:1994eh}. 
Those potentials respect an underlying symmetry called \textit{shape invariance}~\cite{Gendenshtein:1983skv}, which enables one to obtain all the discrete energy eigenvalues by an elementary algebra without any complicated calculation. 
It is also known that any polynomial that satisfies a three-term recurrence relation \textit{and} a second-order ordinary differential equation is restricted to be either the Hermite, Laguerre, or Jacobi polynomial by Bochner's theorem~\cite{1884_Routh, bochner1929}.
The assumptions in the theorem must therefore be loosened to open a new window for non-classical exact solutions in quantum mechanics. 
A way to go beyond the theorem is to apply the Krein--Adler transformation~\cite{ctx59091312420007871, Adler1994} to the classical-orthogonal-polynomial exact solutions. 

In Sections~\ref{subsec:dosc1} and \ref{subsec:dosc2}, the simplest two examples that admit non-classical exact solutions are reviewed based on elementary calculations without employing the Krein--Adler transformation, for readers who are not familiar with it. 
Then, in Section~\ref{subsec:katrf}, the Krein--Adler transformation is briefly summarised, giving a unified understanding to the two cases and offering a way to construct infinitely many exactly solvable potentials both in quantum mechanics and the corresponding Fokker--Planck equation in stochastic inflation. 

\subsection{Quantum deficient oscillator I}
\label{subsec:dosc1}

\begin{figure}
	\centering
	\includegraphics[width = 0.995\linewidth]{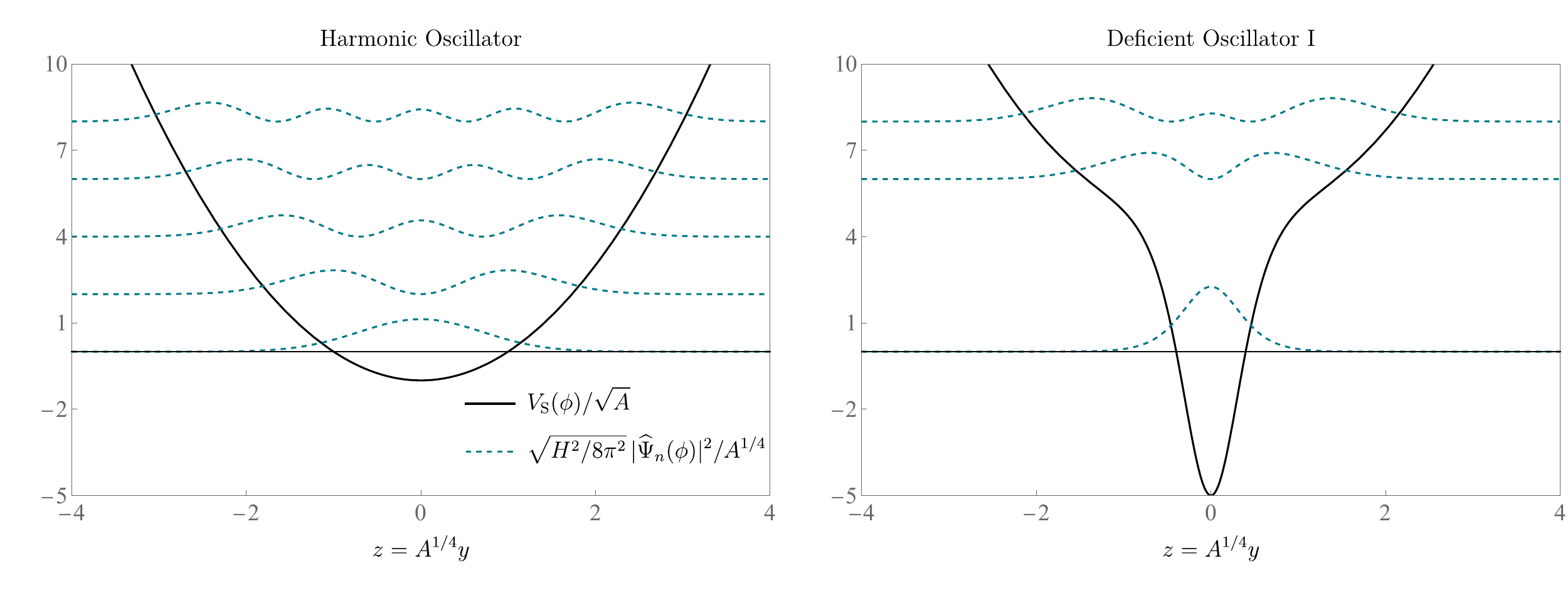}
	\caption{
		The solid curve shows the potential (Eq.~(\ref{eq:c1_pot}) in the right panel), while the dashed curves do the wavefunctions (Eq.~(\ref{eq:df1_wf_nom}) in the right panel) squared. 
		Each wavefunction squared is rescaled by factor two for illustrative purpose, and is displayed with the offset of its energy. 
        It is $\lambda_{n} / \sqrt{A} = 0$ for $n = 0$ and $\lambda_{n} / \sqrt{A} = 2 (n + 2)$ for $n \geq 1$ in the right panel. 
	}
	\label{fig:df1_wfsq}
\end{figure}

Our first exactly solvable model of interest belongs to a class of the ones that can be obtained by a deformation of the quantum harmonic oscillator, as will be seen in Section.~\ref{subsec:katrf} in a general framework under the Krein--Adler transformation. 
The potential in the quantum-mechanical side is given by 
\begin{equation}
	V_{\rm S} (\phi) 
	= \sqrt{A} \, \frac{4 z^{6} + 16 z^{4} + 29 z^{2} - 5}{(2 z^{2} + 1)^{2}} 
	= \sqrt{A} \, \qty[ z^{2} + 3 + 8 \frac{ 2 z^{2} - 1 }{ (2 z^{2} + 1)^{2} } ] 
	\,\, , 
	\qquad 
	z \coloneqq A^{1/4} y 
	\,\, , 
	\qquad 
	A > 0 
	\,\, . 
	\label{eq:c1_pot}
\end{equation}
The variable $z$ has been introduced and called the sinusoidal coordinate in~\cite{Nieto:1978hb, Odake:2006fj} (as well as $z$ in Section~\ref{subsec:hodemo}). 
This can be regarded as an anharmonic oscillator in some sense though the ``perturbed'' term does not have to be small. 
Its series expansion around $z = 0$ is given by $V_{\rm S} (\phi) / \sqrt{A} = -5 + 49 z^{2} + \mathcal{O} (z^{4})$, where the offset ensures that the ground-state energy vanishes (the potential (\ref{eq:c1_pot}) is displayed in the right panel of Figure~\ref{fig:df1_wfsq}). 
The Schr\"{o}dinger equation for the energy eigenvalue and the wavefunction is then given by 
\begin{equation}
	\qty[ 
		- \dv[2]{z} + \frac{4 z^{6} + 16 z^{4} + 29 z^{2} - 5}{(2 z^{2} + 1)^{2}} 
	] \Psi_{n} (\phi) 
	= \frac{ \lambda_{n} }{ \sqrt{A} } \Psi_{n} (\phi) 
	\,\, . 
    \label{eq:df1_eeq}
\end{equation}
The boundary conditions that the wavefunction must satisfy are the natural ones, \textit{i.e.}~$\Psi_{n} (\phi) \to 0$ as $\abs{ \phi } \to \infty$, in the domain $\phi \in (- \infty, \, + \infty)$. 

The function, 
\begin{equation}
    \Psi_{n = 0} (\phi) 
    = \frac{e^{- z^{2} / 2}}{2 z^{2} + 1} 
    \,\, , 
\end{equation}
satisfies Eq.~(\ref{eq:df1_eeq}) with $\lambda_{n = 0} = 0$, which can therefore be confirmed to be the lowest-energy state of the system. 
The shape-invariant arguments cannot be applied to write down the excited energy state with $\lambda_{n > 0} > 0$, which is a major difference of the potential (\ref{eq:c1_pot}) from the conventional shape-invariant ones including the quantum harmonic oscillator. 
Nevertheless, the excited-energy eigenvalues can be found as 
\begin{equation}
	\frac{ \lambda_{n} }{ \sqrt{A} } 
	= 
	\begin{cases}
		0 &\quad (n = 0) \,\, , \\[1.2ex] 
		2 (n + 2) &\quad (n \geq 1) \,\, . 
	\end{cases}
	\label{eq:df1_ene}
\end{equation}
It can be obtained systematically by the Krein--Adler transformation, which will be shown in Section~\ref{subsec:katrf}. 
One can observe that the present spectrum loses the two eigenstates $\lambda = 2 = 2 \cdot 1$ and $\lambda = 4 = 2 \cdot 2$, which are present in the quantum harmonic oscillator, by comparing Eqs.~(\ref{eq:df1_ene}) with (\ref{eq:pre_ex_hoev}). 
The remaining part of the spectrum remains unchanged. 

The function $u_{n} (z)$ is introduced here by factoring the ground-state wavefunction out of the entire function as 
\begin{equation}
	\Psi_{n} (\phi) 
	= \frac{e^{- z^{2} / 2}}{2 z^{2} + 1} u_{n} (z) 
	\,\, . 
	\label{eq:df1_ext}
\end{equation}
This enables us to reduce Eq.~(\ref{eq:df1_eeq}) together with the spectrum (\ref{eq:df1_ene}) to another differential equation with its polynomial solution 
\begin{equation}
	\qty[ 
		\dv[2]{z} - 2 \frac{2 z^{3} + 5 z}{2 z^{2} + 1} \dv{z} + 2 (n + 2) 
	] u_{n} (z) 
	= 0 
	\,\, , 
    \qquad 
    n \geq 1 
    \,\, . 
	\label{eq:c1_diffeq_fz}
\end{equation}
The pedagogical method is employed in the following analysis. 
It does not rely on any sophisticated construction method such as the Krein--Adler transformation introduced in Section~\ref{subsec:katrf}, and a wide range of the readers is hence ready to reach acceptance to the solution that will be non-trivially obtained from now on. 
With $a_{n, \, 0} \neq 0$ or $a_{n, \, 1} \neq 0$, the ansatz, 
\begin{equation}
	u_{n} (z) 
	= \sum_{k = 0}^{\infty} a_{n, \, k} z^{k} 
	\,\, , 
	\label{eq:df1_ser_fb}
\end{equation}
yields the recurrence relations satisfied by the coefficients, $a_{n, \, 2} + (n + 2) a_{n, \, 0} = 0$, $3 a_{n, \, 3} + (n - 3) a_{n, \, 1} = 0$, and 
\begin{equation}
        (k + 4) (k + 3) a_{n, \, k + 4} + 2 ( 
			k^{2} - 2 k + n - 6 
		) a_{n, \, k + 2} 
		+ 4 ( n - k + 2 ) a_{n, \, k} 
        = 0 
		\,\, . \label{eq:df1_inc5} 
\end{equation}
Note that this holds for $k \geq 0$. 
Starting from $a_{n, \, 0}$ and $a_{n, \, 1}$, Eq.~(\ref{eq:df1_inc5}) determines the remaining terms in sequence. 

The second-order recurrence relation (\ref{eq:df1_inc5}) can be solved exactly, 
\begin{subequations}
    \label{eq:df1_ak_gen}
    \begin{align}
        \frac{a_{n, \, k}}{a_{n, \, 0}} 
        &= \frac{(-)^{k/2} 2^{k}}{k!} \qty( \frac{n}{2} + 1 ) \qty[ 
            \frac{n}{2} - k \qty( \frac{k}{2} - 1 ) 
        ] 
        \frac{ \Gamma (n/2) }{ \Gamma \qty[ (n - k)/2 + 2 ] } 
        & 
        \text{for}\quad 
        & 
        k = 0, \, 2, \, \dots 
        \,\, , 
        \label{eq:df1_ak1_gen}
        \\ 
        \frac{a_{n, \, k}}{a_{n, \, 1}} 
        &= \frac{ (-)^{(k-1)/2} 2^{k-1} }{k!} 
        \qty[ 
            \frac{n}{2} - k \qty( \frac{k}{2} - 1 ) 
        ] 
        \frac{ \Gamma \qty[ (n+1)/2 ] }{ \Gamma \qty[ (n-k)/2 + 2 ] } 
        & 
        \text{for}\quad 
        & 
        k = 1, \, 3, \, \dots 
        \,\, . 
        \label{eq:df1_ak2_gen}
    \end{align}
\end{subequations}
Substitution of Eqs.~(\ref{eq:df1_ak_gen}) into Eq.~(\ref{eq:df1_ser_fb}) gives the general solution $u_{n} (z)$ to Eq.~(\ref{eq:c1_diffeq_fz}), but it is not normalisable in general. 
For $k = 2 \ell$ ($\ell = 0, \, 1, \, \dots$), the sequence (\ref{eq:df1_ak1_gen}) terminates at $\ell = n/2 + 2$ when $n$ is even, while for $k = 2 \ell - 1$ ($\ell = 1, \, 2, \, \dots$), the sequence (\ref{eq:df1_ak2_gen}) terminates at $\ell = (n+3) / 2$ when $n$ is odd. 
This implies that, depending on whether $n \geq 1$ is odd or even, the normalisable and hence physical solution is constructed from one of the branches. 
For the even-$n$ branch, the infinite summation under Eq.~(\ref{eq:df1_ak1_gen}) gives the closed-form expression for the wavefunction, given by\footnote{
	Kummer's confluent hypergeometric function~\cite{NIST:DLMF} appears in Eq.~(\ref{eq:df1_fn_form1f1-a}), which is a special case of the generalised hypergeometric function defined by~\cite{Duverney:2024qzj} 
	\begin{equation}
		{}_{p} F_{q} \qty( 
			\begin{matrix} 
				a_{1}, \, \dots, \, a_{p} \\ 
				b_{1}, \, \dots, \, b_{q} 
			\end{matrix}
			~ \middle| ~ z 
		) 
		\coloneqq 
		\sum_{k=0}^{\infty} \frac{ 
			(a_{1})_{k} \cdots (a_{p})_{k} 
		}{ 
			(b_{1})_{k} \cdots (b_{q})_{k} 
		} \frac{z^{k}}{k!} 
		\,\, . 
	\end{equation}
	The Pochhammer symbol, $(a)_{n} \coloneqq \Gamma (a + n) / \Gamma (a)$, has been introduced. 
}
\begin{equation}
	u_{n} (z) 
    = a_{n, \, 0} 
    \qty[ 
		{}_{1} F_{1}  
		\qty( 
			\begin{matrix}
                			-n/2 \\ 3/2 
            		\end{matrix}
            		~ \middle| ~ z^{2} 
		) 
        - \frac{2}{3} (n + 3) z^{2} \cdot {}_{1} F_{1} \qty( 
			\begin{matrix} 
				- n/2 \\ 5/2 
			\end{matrix}
			~ \middle| ~ z^{2} 
		) 
		- \frac{4}{3} (n + 2) z^{4} \cdot {}_{1} F_{1} \qty( 
			\begin{matrix}
				- n/2 + 1 \\ 5/2 
			\end{matrix}
			~ \middle| ~ z^{2} 
		) 
	] 
	\,\, , 
	\label{eq:df1_fn_form1f1-a}
\end{equation} 
for $n = 2, \, 4, \, \dots$. 
On the other hand, the closed-form result of the infinite summation that follows from Eq.~(\ref{eq:df1_ak2_gen}) reads 
\begin{equation}
	u_{n} (z) 
	= a_{n, \, 1} \, z \, 
	\qty[ 
		(2 z^{2} + 1) \cdot {}_{1} F_{1} \qty( 
			\begin{matrix}
				- (n+1)/2 \\ 3/2 
			\end{matrix}
			~ \middle| ~ z^{2} 
		) 
		+ \frac{2}{3} z^{2} (2 z^{2} - 1) \cdot {}_{1} F_{1} \qty( 
			\begin{matrix}
				- (n-1)/2 \\ 5/2
			\end{matrix}
			~ \middle| ~ z^{2} 
		) 
	] 
    \,\, , 
    \label{eq:df1_fn_form1f1-b}
\end{equation}
for $n = 1, \, 3, \, \dots$. 
Since Eqs.~(\ref{eq:df1_fn_form1f1-a}) and (\ref{eq:df1_fn_form1f1-b}) turn out to be the same functional forms as will be seen shortly, the former will be focussed on in what follows. 

Let us determine $a_{n, \, 0}$ in such a way that the coefficient of the highest-order term in $u_{n} (z)$ becomes unity, although this procedure here is not really necessary since the overall constant in the wavefunction will be determined to satisfy the normalisation condition to ensure the standard probability interpretation. 
The function $u_{n} (z)$ involves several hypergeometric functions and so seems complicated, for normalisation for instance. 
However, by using some identities such as~\cite{Duverney:2024qzj} 
\begin{equation}
	(c - a) z \cdot {}_{1} F_{1} \qty( \begin{matrix} a \\ c + 1 \end{matrix} ~\middle| ~ z) + c (c-1) \cdot {}_{1} F_{1} \qty( \begin{matrix} a \\ c - 1 \end{matrix} ~\middle| ~ z) = c (c - 1 + z) \cdot {}_{1} F_{1} \qty( \begin{matrix} ~a~ \\ ~c~ \end{matrix} ~\middle| ~ z) 
    \,\, , 
    \label{eq:df1_1f1id}
\end{equation}
and the relations between ${}_{1} F_{1}$ and the Hermite polynomials~\cite{Duverney:2024qzj}, 
\begin{subequations}
	\begin{align}
		H_{n} (z) 
		&= (-)^{n/2} \frac{ \Gamma (n + 1) }{ \Gamma (n/2 + 1) } \cdot {}_{1} F_{1} \qty( \begin{matrix} - n / 2 \\  1/2 \end{matrix} ~\middle| ~ z^{2} ) 
        &\quad \text{for even} ~~ n 
		\,\, , 
		\\ 
		H_{n+1} (z) 
		&= (-)^{n/2} \frac{ \Gamma (n + 2) }{ \Gamma (n/2 + 1) } \cdot 2 z \cdot {}_{1} F_{1} \qty( \begin{matrix} - n / 2 \\ 3 / 2 \end{matrix} ~\middle| ~ z^{2} ) 
        &\quad \text{for odd} ~~ n 
		\,\, , 
	\end{align}
    \label{eq:df1_herm}
\end{subequations}
it turns out that Eq.~(\ref{eq:df1_fn_form1f1-a}) can be reexpressed in terms solely of the Hermite polynomials as 
\begin{equation}
	\widehat{u}_{n} (z) 
	= \frac{1}{(n+1) 2^{n+1}} \qty[ 
		2 \frac{n + 2}{n} \qty( z^{3} - \frac{z}{2} ) H_{n + 1} (z) 
		- \qty( \frac{n + 2}{n} z^{2} + \frac{1}{2} ) H_{n + 2} (z)  
	] 
	\,\, , 
	\qquad 
	n \geq 1 
	\,\, . 
    \label{eq:df1_sersol}
\end{equation}
Note that $\widehat{u}_{n} (z)$ represents that the function $u_{n} (z)$ is normalised in the sense that $a_{n, \, 0}$ is fixed so that the highest-order coefficient is unity as was announced above. 
The odd-$k$ solution can also be reduced to the exactly same expression as Eq.~(\ref{eq:df1_sersol}) though a bit more complicated manipulations are needed. 
The inner product among the polynomials $u_{n} (z)$ reads 
\begin{equation}
	\int_{\mathbb{R}} \dd z \, \frac{e^{-z^2}}{(2 z^{2} + 1)^{2}} [ \, \widehat{u}_{n} (z) ]^{2} 
	= \frac{n+2}{8} \Gamma \qty( \frac{n}{2} ) \Gamma \qty( \frac{n+2}{2} ) 
	\,\, , 
    \qquad 
    n \geq 1 
    \,\, . 
    \label{eq:df1_orthof}
\end{equation}
The weight function, $w (z) = e^{- z^{2}} / (2 z^{2} + 1)^{2}$, is sometimes called the perturbed Gaussian weight in the literature~\cite{10.1063/5.0138122}. 

The excited-state wavefunction is written as 
\begin{equation}
	\widehat{\Psi}_{n} (\phi) 
	= \mathsf{N}_{n} \frac{e^{- z^{2} / 2}}{2 z^{2} + 1} \widehat{u}_{n} (z) 
	\,\, , 
	\qquad 
	n \geq 1 
	\,\, , 
\end{equation}
introducing the normalisation constant $\mathsf{N}_{n}$. 
The normalisation condition imposed on the wavefunction is given by 
\begin{equation}
	1 
	= \int_{\mathbb{R}} \dd \phi \, [ \widehat{ \Psi }_{n} (\phi) ]^{2} 
	= \frac{\mathsf{N}_{n}^{2}}{A^{1/4}} \sqrt{ \frac{H^{2}}{8 \pi^{2}} } \, \int \dd z \, \frac{e^{- z^{2}}}{(2 z^{2} + 1)^{2}} [ \, \widehat{u}_{n} (z) ]^{2} 
	\,\, , 
    \label{eq:df1_orthowf}
\end{equation}
where in the last and second-to-last expressions the integrands should be understood as those written in terms of $z$ and $\phi$, respectively, through the relation $z = A^{1/4} y = A^{1/4} (\sqrt{8 \pi^{2} / H^{2}}) \phi$. 
Together with the relation (\ref{eq:df1_orthof}), 
Eq.~(\ref{eq:df1_orthowf}) determines $\mathsf{N}_{n}$. 
The normalised ground-state wavefunction can thus be obtained from Eq.~(\ref{eq:df1_ext}). 
To summarise, one arrives at 
\begin{equation} 
    \widehat{\Psi}_{n} (\phi) 
    = \qty( \sqrt{A} \, \frac{8 \pi^{2}}{H^{2}} )^{1/4} \frac{ e^{- z^{2}/2} }{2 z^{2} + 1} \times 
    \left\{ ~ 
    \begin{aligned}
        & \frac{ \sqrt{2} }{\pi^{1/4}} 
            && (n = 0) \,\, , 
        \\
        &  
	\frac{1}{(n + 1) 2^{n+1}}  
	\qty[ \frac{n+2}{8} \Gamma \qty( \frac{n}{2} ) \Gamma \qty( \frac{n+1}{2} ) ]^{- 1/2} 
            \\
            & \quad \times 
	\qty[ 
		\qty( \frac{n+2}{n} z^{2} + \frac{1}{2} ) H_{n + 2} (z) 
		- 2 \frac{n+2}{n} \qty( z^{3} - \frac{z}{2} ) H_{n + 1} (z) 
	] 
            && (n \geq 1) \,\, . 
    \end{aligned}
    \right. 
    \label{eq:df1_wf_nom}
\end{equation}

The right panel of Figure~\ref{fig:df1_wfsq} shows the wavefunction (\ref{eq:df1_wf_nom}) squared as well as the potential (\ref{eq:c1_pot}), see also those of the quantum harmonic oscillator (\ref{eq:pre_ex_hopot}) in the left panel for comparison. 
As can be seen, the energy spectrum is shared by each other except the first and second excited states. 
Due to the larger mass around the origin in the quantum deficient oscillator, there is no eigenstates between $\lambda_{n = 0} = 0$ and $\lambda_{n = 1} = 6$. 
It will be seen in Section~\ref{sec:statds} that the analytical wavefunctions (\ref{eq:df1_wf_nom}) together with the spectrum (\ref{eq:df1_ene}) give rise to the corresponding exact solution to the Fokker--Planck equation, which describes the dynamics of the scalar field in de Sitter spacetime. 

\subsection{Quantum deficient oscillator II}
\label{subsec:dosc2}

\begin{figure}
	\centering
	\includegraphics[width = 0.995\linewidth]{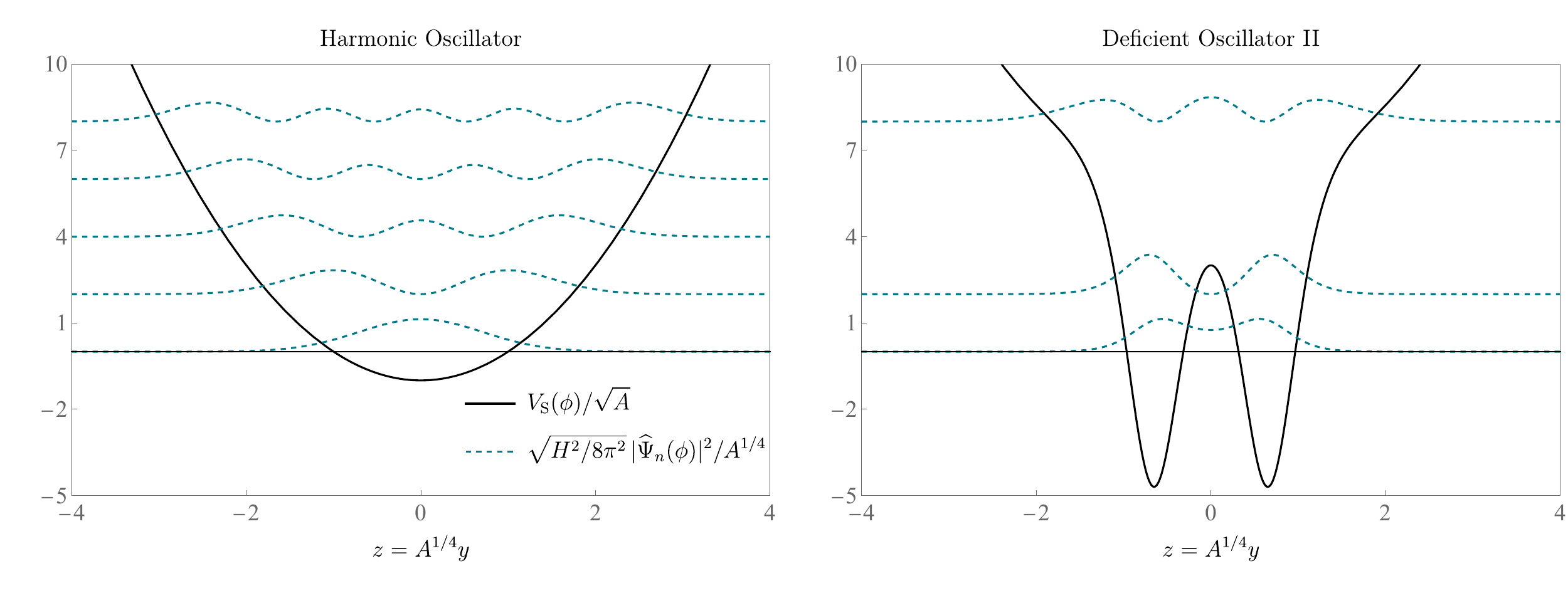}
	\caption{
		In the right panel, the solid curve shows the potential (\ref{eq:df2_pot}), while the dashed curves do the wavefunction (\ref{eq:df2_wf_nom}) squared. 
		Each wavefunction is displayed with the offset of its energy, $\lambda_{n} / \sqrt{A} = 0$ for $n = 0$ and $\lambda_{n} / \sqrt{A} = 2 (n + 2)$ for $n \geq 1$. 
	}
	\label{fig:df2_wfsq}
\end{figure}

The model discussed in Sec.~\ref{subsec:dosc1} is the simplest but non-classical one that shares its energy spectrum with the quantum harmonic oscillator, except the first two excited states in the latter. 
It demonstrates non-trivial acquisition of the exact eigenvalues and eigenfunctions, for the complicated but still the single-well potential given by Eq.~(\ref{eq:c1_pot}). 
Let us now move on to the second model, a double-well potential that admits to an analytical solution, 
\begin{equation}
	V_{\rm S} (\phi) 
	= \sqrt{A} \, \frac{ 
		16 z^{10} + 48 z^{8} + 152 z^{6} + 72 z^{4} - 279 z^{2} + 27 
	}{ 
		(4 z^{4} + 3)^{2} 
	} 
	= \sqrt{A} \, \qty[ z^{2} + 3 + 32 \frac{ 4 z^{6} - 9 z^{2} }{ (4 z^{4} + 3)^{2} } ] 
	\,\, . 
	\label{eq:df2_pot}
\end{equation}
The variable $z$ is related to $y$ in the same way as in the previous case. 
The series expansion of this potential around $z = 0$ is given by $V_{\rm S} (\phi) / \sqrt{A} = 3 - 31 z^{2} + (896 / 9) z^{6} + \mathcal{O} (z^{10})$, where the second term leads to a local maximum, indicating the unstable point for $z=0$
(the potential (\ref{eq:df2_pot}) is displayed in the right panel of Figure~\ref{fig:df2_wfsq}).
The Schr\"{o}dinger equation reads 
\begin{equation}
	\qty[ 
		- \dv[2]{z} + \frac{ 
		16 z^{10} + 48 z^{8} + 152 z^{6} + 72 z^{4} - 279 z^{2} + 27 
	}{ 
		(4 z^{4} + 3)^{2} 
	}  
	] \Psi_{n} (\phi) 
	= \frac{ \lambda_{n} }{ \sqrt{A} } \Psi_{n} (\phi) 
	\,\, . 
    \label{eq:df2_eeq}
\end{equation}

The energy spectrum is identified to be 
\begin{equation}
	\frac{ \lambda_{n} }{ \sqrt{A} } 
	= 
	\begin{cases}
		0 &\quad (n = 0) \,\, , \\[1.2ex] 
		2 &\quad (n = 1) \,\, , \\[1.2ex] 
		2 (n + 2) &\quad (n \geq 2) \,\, . 
	\end{cases}
\end{equation}
The factoring procedure such as Eq.~(\ref{eq:df1_ext}) is now performed for $n \geq 2$, that is, 
\begin{equation}
	\Psi_{n} (\phi) 
	= \frac{e^{-z^{2}/2}}{4 z^{4} + 3} u_{n} (z) 
	\,\, , 
    \qquad 
    n \geq 2 
    \,\, . 
\end{equation}
Rewriting Eq.~(\ref{eq:df2_eeq}) in terms of $u_{n} (z)$, substitution of the series-solution ansatz (\ref{eq:df1_ser_fb}) gives the recurrence relations, 
\begin{align}
	0 
	&= 3 (k + 6) (k + 5) a_{n, \, k + 6} 
	+ 6 \qty[ n - (k + 4) ] a_{n, \, k + 4} 
    \notag \\ 
	&\quad + 4 \qty[ 
		12 - 8 (k + 2) + (k + 2) (k + 1) 
	] a_{n, \, k + 2} 
	+ 8 \qty[ (n + 4) - k ] a_{n, \, k} 
    \label{eq:df2_rec}
\end{align}
for $k = 0, \, 1, \, \dots$. 
The third-order recurrence relation (\ref{eq:df2_rec}) can be solved non-trivially and exactly again. 
Due to the same structure as the previous case, let us focus on the branch including $a_{n, \, 0}$, to which the general expression is given by 
\begin{equation}
	\frac{a_{n, \, k}}{a_{n, \, 0}} 
	= \frac{ (-4)^{k / 2} }{ \Gamma (k + 1) } 
	\frac{(n/2 + 1) \Gamma (n/2)}{ \Gamma \qty[ n/2 - (k/2 - 3) ] } 
	\qty{ 
		\frac{n}{2} \qty[ \frac{n}{2} - \qty( k - 2 ) ] + \frac{2 k}{3} \qty( \frac{k}{2} - 1 ) \qty( \frac{k}{2} - 2 )^{2} 
	} 
	\,\, , 
    \label{eq:df2_ak_gen}
\end{equation}
for $k = 0, \, 2, \, 4, \, \dots$. 
The coefficient $a_{n, \, 0}$ is determined so as to the leading coefficient of $u_{n} (z)$ becomes unity, that is, $a_{n, \, 0} = (-1/4)^{n/2} \cdot (3/2) \cdot [ (n + 1)/(n - 1) ] \cdot \Gamma (n) / \Gamma (n/2)$. 
The infinite summation (\ref{eq:df1_ser_fb}) under Eq.~(\ref{eq:df2_ak_gen}) can again be written in a closed form. 
For instance, it can be expressed in terms of several hypergeometric functions as
\begin{align}
    u_{n} (z) 
    &= \frac{2}{3} a_{n, \, 0} \left[
        \frac{3}{2} \cdot {}_{1} F_{1} \qty( 
            \begin{matrix} - n / 2 - 2 \\ 1 / 2 \end{matrix} ~ \middle| ~ z^{2} 
        ) 
        + 6 z^{2} \cdot {}_{1} F_{1} \qty( 
            \begin{matrix} - n / 2 - 2 \\ 3 / 2 \end{matrix} ~ \middle| ~ z^{2} 
        ) 
    \right. 
    \notag \\ 
    &\quad \left. 
        - \frac{16}{15} \qty( \frac{n}{2} + 1) \, z^{6} \cdot {}_{1} F_{1} \qty( 
            \begin{matrix} - n / 2 + 1 \\ 7 / 2 \end{matrix} ~ \middle| ~ z^{2} 
        ) 
        + \frac{32}{105} \qty( \frac{n^{2}}{4} - 1 ) \, z^{8} \cdot {}_{1} F_{1} \qty( 
            \begin{matrix} - n / 2 + 2 \\ 9 / 2 \end{matrix} ~ \middle| ~ z^{2} 
        )  
    \right] 
    \,\, , 
    \qquad 
    n \geq 2 
    \,\, . 
\end{align}
Using the relations (\ref{eq:df1_1f1id}) and (\ref{eq:df1_herm}), it turns out that $u_{n} (z)$ for $n \geq 2$ can be expressed in terms of several Hermite polynomials, as 
\begin{equation}
	\widehat{u}_{n} (z) 
	= \frac{ 
		(n+1) (n+2) (4 z^{4} + 3) H_{n} (z) - 8 (n + 2) z^{3} H_{n + 1} (z) + 3 (2 z^{2} + 1) H_{n + 2} (z) 
	}{ 
		2^{n+2} n (n-1) 
	} 
    \,\, . 
    \label{eq:df2_sersol}
\end{equation}
The inner product among the polynomials (\ref{eq:df2_sersol}) can be calculated to give 
\begin{equation}
	\int_{\mathbb{R}} \dd z \, \frac{e^{-z^{2}}}{(4 z^{4} + 3)^{2}} [ \widehat{u}_{n} (z) ]^{2} 
	= \frac{\sqrt{\pi}}{2^{n+4}} (n + 1) (n + 2) \Gamma (n-1) 
	\,\, , 
	\qquad 
	n \geq 2 
	\,\, . 
    \label{eq:df2_orthof}
\end{equation}

The normalisation of the wavefunction is determined by using Eq.~(\ref{eq:df2_orthof}) to arrive at 
\begin{equation}\label{eq:df2_wf_nom}
	\widehat{ \Psi }_{n} (\phi) 
	= \qty( \frac{\sqrt{A}}{\pi} \frac{8 \pi^{2}}{H^{2}} )^{1/4} \frac{ e^{- z^{2} / 2} }{ 4 z^{4} + 3 } 
    \times 
	\begin{cases}
		\displaystyle 2 \sqrt{6} \qty( z^{2} + \frac{1}{2} ) 
		&\quad (n = 0) \,\, , \\[2.0ex] 
		\displaystyle 4 \qty( z^{3} + \frac{3}{2} z ) 
		&\quad (n = 1) \,\, , \\[2.0ex] 
		\displaystyle \frac{4 \cdot 2^{n/2}}{ \sqrt{ (n + 1) (n + 2) \Gamma (n - 1) } } \, \widehat{u}_{n} (z) 
		&\quad (n \geq 2) 
		\,\, . 
	\end{cases}
\end{equation}
The right panel of Figure~\ref{fig:df2_wfsq} shows the normalised wavefunctions squared. 
Note that, as in Figure~\ref{fig:df1_wfsq}, the wavefunctions squared are rescaled by factor two for illustrative purpose. 
Compared to the quantum harmonic oscillator, it can be observed that the second and third excited states are deficient in the transformed oscillator, Eq.~(\ref{eq:df2_pot}). 

\subsection{Deficient oscillators from Krein--Adler transformation}
\label{subsec:katrf}

\begin{figure}
    \centering
    \includegraphics[width = 0.8\linewidth]{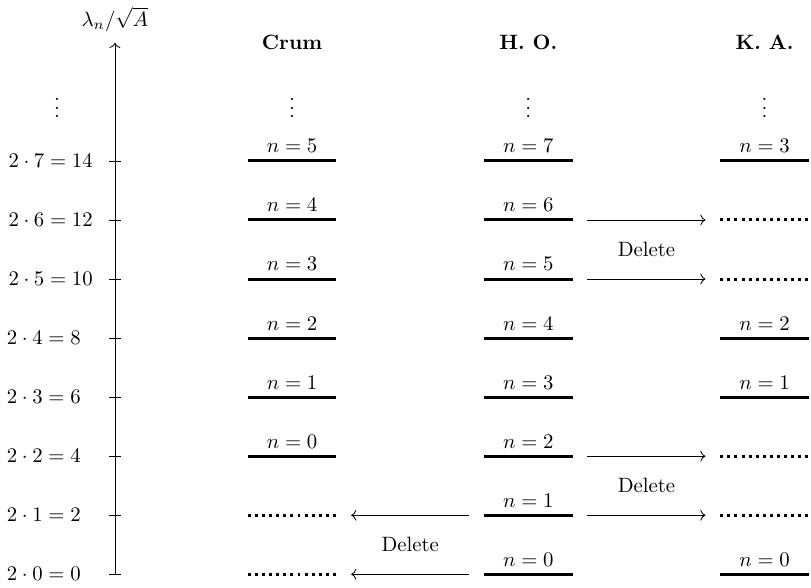}
    \caption{
        Typical energy spectra of the quantum harmonic oscillator, the Crum-transformed, and the Krein--Adler-transformed oscillators. 
        The dotted horizontal line indicates that the eigenstate is absent in each transformed system. 
    }
    \label{fig:rem}
\end{figure}

The fact that the two models discussed in Sections~\ref{subsec:dosc1} and \ref{subsec:dosc2} can non-trivially and exactly be solved itself is attractive, but one may wonder what the structure is behind these models that enables the systems to be solved analytically, and whether there is a unified framework that simultaneously explains those two models. 
Our first model (\ref{eq:c1_pot}) is the simplest case in several previous works by Hongler and his coauthor~\cite{1982PhLA...91..396H, HZ1982}, where the analytical solution was obtained directly by solving the corresponding Fokker--Planck equation. 
It was later pointed out in Ref.~\cite{Junker:1998sj} that Hongler's solution corresponds to the eigensystem of the deformed quantum anharmonic oscillator constructed by Junker and Roy. 
It is well known that particular choices of parameters in their deformation are equivalent to the simplest cases of the Krein--Adler-transformed anharmonic oscillator. 

In general, the Krein--Adler transformation can systematically generate infinitely many transformed anharmonic oscillators that are not in the class of Junker--Roy's deformation. 
As will be seen below, the other model~(\ref{eq:df2_pot}) considered in this article is the second simplest model of the transformation. 
This is why the Krein--Adler transformation is briefly recapitulated here, giving a unified understanding for the two potentials discussed in the previous subsections, and presenting the general formula for the potentials and wavefunctions that will be used in Section~\ref{sec:statds}. 

Starting from the quantum harmonic oscillator, $V_{\rm HO} (\phi)$ in Eq.~(\ref{eq:pre_ex_hopot}), the transformed one in the present context is given by 
\begin{equation}
	V_{\mathscr{D}} (\phi) 
	\coloneqq V_{\rm HO} (\phi) - 2 \qty( \frac{H^{2}}{8 \pi^{2}} ) \dv[2]{\phi} \ln \mathrm{W} [ 
		\widehat{\Psi}_{d_{1}} (\phi), \, \widehat{\Psi}_{d_{2}} (\phi), \, \cdots, \, \widehat{\Psi}_{d_{\ell}} (\phi)
	] 
	\,\, . 
	\label{eq:cat_def}
\end{equation}
Note that the potential (\ref{eq:cat_def}) is the one in the Schr\"{o}dinger equation; it is not a potential in the Fokker--Planck equation. 
The function $\mathrm{W}$ appearing in Eq.~(\ref{eq:cat_def}) is the Wronskian, which is, in general, defined by 
\begin{equation}
	\mathrm{W} [ 
		p_{1} (z), \, p_{2} (z), \, \dots, \, p_{n} (z) 
	] 
	\coloneqq \det \mqty[ 
		p_{1} (z) & p_{2} (z) & \cdots & p_{n} (z) \\[1.5ex] 
		\displaystyle \dv{p_{1} (z)}{z} & \displaystyle \dv{p_{2} (z)}{z} & \cdots & \displaystyle \dv{p_{n} (z)}{z} \\[1.5ex] 
		\vdots & \vdots & \ddots & \vdots \\[1.5ex] 
		\displaystyle \dv[n-1]{p_{1} (z)}{z} & \displaystyle \dv[n-1]{p_{2} (z)}{z} & \cdots & \displaystyle \dv[n-1]{p_{n} (z)}{z} 
	] 
	= \det \qty[ 
		\dv[i-1]{p_{j} (z)}{z} 
	]_{i, \, j} 
\end{equation}
for $i, \, j = 1, \, 2, \, \cdots, \, n$. 
In Eq.~(\ref{eq:cat_def}), the variable to differentiate the wavefunctions $\widehat{\Psi}_{d_{k}} (\phi)$ can be $\phi$, $y$, or $z$, since they are proportional to each other and the overall coefficient is irrelevant after taking logarithmic differentiation. 
The subscript $\mathscr{D}$ denotes a set of the energy levels present in $V_{\rm HO}$ but absent in the newly constructed model, $\mathscr{D} \coloneqq \{ d_{1}, \, d_{2}, \, \dots, \, d_{\ell} \}$. 
It should be chosen in such a way that the members in $\mathscr{D}$ satisfy 
\begin{equation}
    \forall n \in \mathbb{Z}_{\geq 0} ~,\quad
    \prod_{k=1}^{\ell} (n-d_k) \geq 0 ~,
\end{equation}
so that the newly constructed model has no singularities in the domain, $\phi \in (- \infty, \, + \infty)$.
This condition implies that $\mathscr{D}$ must be either (a) a set of several pairs of two successive non-negative integers with, consequently, even $\ell$, or (b) a set of $\ell$ successive integers from $0$.
The latter, (b), corresponds to Crum's transformation~\cite{10.1093/qmath/6.1.121}, which is usually discussed in the context of exactly solvable quantum mechanics. 
See also Figure~\ref{fig:rem}. 

By the Krein--Adler transformation, the wavefunctions $\widehat{\Psi}_{n} (\phi)$ are mapped according to 
\begin{equation}
    \widehat{\Psi}_{n} (\phi)
    \mapsto \frac{ 
		\mathrm{W} [ 
			\widehat{\Psi}_{d_{1}} (\phi), \, 
			\widehat{\Psi}_{d_{2}} (\phi), \, 
			\cdots, \, 
			\widehat{\Psi}_{d_{\ell}} (\phi), \, 
			\widehat{\Psi}_{n} (\phi) 
		] 
	}{ 
		\mathrm{W} [ 
			\widehat{\Psi}_{d_{1}} (\phi), \, 
			\widehat{\Psi}_{d_{2}} (\phi), \, 
			\cdots, \, 
			\widehat{\Psi}_{d_{\ell}} (\phi) 
		] 
	} 
	\,\, . 
\end{equation}
When $n \in \mathscr{D}$, the numerator vanishes, and the states are therefore \textit{deleted} by the transformation.
It is then convenient to introduce the notation $\check{n} = \check{n} (n)$, where $\check{n}$ is in general a function of $n$ and reflects the deletion of several energy levels specified by the set $\mathscr{D}$, $\check{n} \in \mathbb{Z}_{\geq 0} \backslash \mathscr{D}$. 
The normalised eigenfunctions of the Hamiltonian with the potential \eqref{eq:cat_def} read 
\begin{equation}
    \widehat{\Psi}_{\mathscr{D}, \, n} (\phi) 
    = \left( 2^{\ell} \prod_{k=1}^{\ell} (\check{n} - d_k) \right)^{-\frac{1}{2}} 
    \frac{ 
		\mathrm{W} [ 
			\widehat{\Psi}_{d_{1}} (\phi), \, 
			\widehat{\Psi}_{d_{2}} (\phi), \, 
			\cdots, \, 
			\widehat{\Psi}_{d_{\ell}} (\phi), \, 
			\widehat{\Psi}_{\check{n}} (\phi) 
		] 
	}{ 
		\mathrm{W} [ 
			\widehat{\Psi}_{d_{1}} (\phi), \, 
			\widehat{\Psi}_{d_{2}} (\phi), \, 
			\cdots, \, 
			\widehat{\Psi}_{d_{\ell}} (\phi) 
		] 
	} 
	\,\, ,
    \qquad 
	\check{n} = \check{n} (n) 
	\,\, , 
    \label{eq:katrf_wfgen}
\end{equation}
and the corresponding eigenvalues are 
\begin{equation}
    \frac{\lambda_{n}}{\sqrt{A}} = 2\check{n}
    \,\, . 
\end{equation}

A particular case where $\ell = 2$, or equivalently $\mathscr{D} = \qty{ d_{1}, \, d_{2} = d_{1} + 1 }$, is focussed on in what follows, to demonstrate the Krein--Adler transformation and to reproduce the quantities considered in the previous subsections. 
In such cases, the corresponding wavefunctions of the newly constructed quantum deficient oscillator are given, as a special case of Eq.~(\ref{eq:katrf_wfgen}), by 
\begin{equation}
	\widehat{\Psi}_{\mathscr{D}, \, n} (\phi) 
	= \frac{1}{\sqrt{4 (\check{n} - d_{1}) (\check{n} - d_{2})}} 
    \frac{ 
		\mathrm{W} [ 
			\widehat{\Psi}_{d_{1}} (\phi), \, 
			\widehat{\Psi}_{d_{2}} (\phi), \, 
			\widehat{\Psi}_{\check{n}} (\phi) 
		] 
	}{ 
		\mathrm{W} [ 
			\widehat{\Psi}_{d_{1}} (\phi), \, 
			\widehat{\Psi}_{d_{2}} (\phi) 
		] 
	} 
	\,\, , 
	\qquad 
	\check{n} = \check{n} (n) 
	\,\, . 
    \label{eq:cat_kwfnom}
\end{equation}
The concrete expression of $\check{n}$ depends on the members in $\mathscr{D}$, see below. 
Using this scheme of the Krein--Adler transformation, the exact solutions presented in the previous subsections can be constructed systematically. 

\paragraph*{\textit{Deficient oscillator I}.}
The quantum deficient oscillator solved in Section~\ref{subsec:dosc1} has the energy spectrum (\ref{eq:df1_ene}), in which the first and second excited states of the quantum harmonic oscillator are absent. 
Indeed, the potential (\ref{eq:c1_pot}) is obtained by a the Krein--Adler transformation with those energy levels being deleted by setting $d_1=1$, as 
\begin{equation}
	V_{\mathscr{D} = \{ 1, \, 2 \}} (\phi) 
	= V_{\rm HO} (\phi) - 2 \qty( \frac{H^{2}}{8 \pi^{2}} ) \dv[2]{\phi} \ln \mathrm{W} [ 
		\widehat{\Psi}_{1} (\phi), \, \widehat{\Psi}_{2} (\phi) 
	] 
	= \sqrt{A} \, \frac{ 4 z^{6} + 16 z^{4} + 29 z^{2} - 5 }{ (2 z^{2} + 1)^{2} } 
	\,\, , 
    \label{eq:katrf_c1pot}
\end{equation}
which exactly matches Eq.~(\ref{eq:c1_pot}). 
The wavefunctions can also be obtained by the Krein--Adler scheme as 
\begin{equation}
	\widehat{\Psi}_{\mathscr{D} = \{ 1, \, 2 \}, \, n} (\phi) 
	= \frac{1}{\sqrt{4 (\check{n} - 1) (\check{n} - 2)}} 
	\frac{ 
		\mathrm{W} [ \widehat{\Psi}_{1} (\phi), \, \widehat{\Psi}_{2} (\phi), \, \widehat{\Psi}_{\check{n}} (\phi) ] 
	}{ 
		\mathrm{W} [ \widehat{\Psi}_{1} (\phi), \, \widehat{\Psi}_{2} (\phi) ] 
	} 
	\,\, , 
	\qquad 
	\check{n} 
	\coloneqq 
	\begin{cases}
		~ n &\quad (n = 0) \,\, , \\[1.5ex] 
		~ n + 2 &\quad (n \geq 1) \,\, . 
	\end{cases}
\end{equation}
For the transformed model (\ref{eq:katrf_c1pot}), the energy eigenvalue of the ground state remains the same as that of the original quantum harmonic oscillator. 
On the other hand, those of the first and second excited states in the original system are absent in the transformed model, so, for instance, the first excited state in the new model corresponds to the third excited state, and so on. 
This leads to the concrete expression for $\check{n}$ in terms of $n$. 

\paragraph*{\textit{Deficient oscillator II}.}
In the model considered in Section~\ref{subsec:dosc2}, the second and third energy levels, \textit{i.e.}~$d_{1} = 2$ and $d_{2} = 3$, are deficient. 
In a similar manner to Eq.~(\ref{eq:katrf_c1pot}), the Krein--Adler construction of the potential therefore reads 
\begin{equation}
	V_{\mathscr{D} = \{ 2, \, 3 \}} (\phi) 
	= V_{\rm HO} (\phi) - 2 \qty( \frac{H^{2}}{8 \pi^{2}} ) \dv[2]{\phi} \ln \mathrm{W} [ 
		\widehat{\Psi}_{2} (\phi), \, \widehat{\Psi}_{3} (\phi) 
	] 
	= \sqrt{A} \, \qty[ z^{2} + 3 + 32 \frac{ 4 z^{6} - 9 z^{2} }{ (4 z^{4} + 3)^{2} } ] 
	\,\, , 
\end{equation}
which exactly matches Eq.~(\ref{eq:df2_pot}). 
The wavefunctions and the concrete functional form of $\check{n} (n)$ are given by 
\begin{equation}
	\widehat{\Psi}_{\mathscr{D} = \{ 2, \, 3 \}, \, n} (\phi) 
	= \frac{1}{\sqrt{4 (\check{n} - 2) (\check{n} - 3)}} \frac{ 
		\mathrm{W} [ \widehat{\Psi}_{2} (\phi), \, \widehat{\Psi}_{3} (\phi), \, \widehat{\Psi}_{\check{n}} (\phi) ] 
	}{ 
		\mathrm{W} [ \widehat{\Psi}_{2} (\phi), \, \widehat{\Psi}_{3} (\phi) ] 
	} 
	\,\, , 
	\qquad 
    \check{n} 
	\coloneqq 
	\begin{cases}
		~ n &\quad (n = 0, \, 1) \,\, , \\[1.5ex] 
		~ n + 2 &\quad (n \geq 2) \,\, . 
	\end{cases}
\end{equation}

Before closing this section, we note that further anharmonic, quantum deficient, oscillators can be constructed in this course by taking $d_{1} = 3, \, 4, \, \ldots~$, 
and/or by considering $\ell \neq 2$ cases. 
We shall return to this point later in Section \ref{subsec:gallery}.
It is also worth emphasising here that the choice of $d_1=0$ does not produce any new potential, $V_{\mathscr{D} = \{ 0, \, 1 \}} (\phi) = V_{\rm HO} (\phi) + 4 \sqrt{A}$, $V_{\mathscr{D} = \{ 0, \, 1, \, 2 \}} (\phi) = V_{\rm HO} (\phi) + 6 \sqrt{A}$, and in general $V_{\mathscr{D} = \{ 0, \, 1, \, \dots, \, \ell \}} (\phi) = V_{\rm HO} (\phi) + 2 (\ell + 1) \sqrt{A}$, for the quantum harmonic oscillator respects shape invariance. 

\section{Exact PDFs and statistical moments in de Sitter spacetime}
\label{sec:statds}

Returning to stochastic inflation, the corresponding solution to our Fokker--Planck equation (\ref{eq:pre_fp}) will be studied from now on. 
The analytical distribution functions are constructed from the wavefunctions obtained in the previous section, which enables us to give the analytical expression of the statistical moments and to analytically describe the scalar-field dynamics in de Sitter spacetime. 
At the classical level (in the sense that the corresponding quantum-mechanical system is solved in terms of the classical orthogonal polynomials), there are several situations where the statistical quantities such as the distribution and correlation functions can be obtained exactly. 
The most famous case is a scalar field in the purely quadratic potential, for which the solution of the Fokker--Planck equation is given by a decaying Gaussian function, as reviewed in Section~\ref{subsec:hodemo}. 
The corresponding quantum-mechanical system is the harmonic oscillator, and thus also exactly solvable in terms of the Hermite polynomials. 

A class of exact solutions in stochastic inflation in terms of the classical orthogonal polynomials, focusing on the shape-invariant systems in quantum mechanics, is summarised in Ref.~\cite{Honda:2024evc}. 
In particular, the four potentials given in Ref.~\cite{Honda:2024evc} exhaust the setups where the corresponding quantum-mechanical systems are solved in terms of the classical orthogonal polynomials and with only \textit{discrete} eigenstates. 
One therefore needs to consider other setups where both discrete and continuous eigenstates are present, or to consider other setups where shape invariance can no longer be found.  
The latter option has been chosen in this article, which is why in the previous section the Krein--Adler-transformed quantum deficient oscillators were considered. 

Let us start with the spectral decomposition of the solution to the Fokker--Planck equation, given by (\ref{eq:pre_sdcmp}). 
Given that it is linear in its solution, considering the Dirac-$\delta$ initial distribution~(\ref{eq:initial_dist}) is sufficient as is implemented in Section~\ref{subsec:dfsh},
\begin{equation}
	f (\phi, \, N = N_{0}) 
	= \delta_{\rm D} (\phi - \phi_{0}) 
	\,\, . 
    \label{eq:exdist_coni}
\end{equation}
This is because the solution to the concentrated initial distribution (\ref{eq:exdist_coni}) can be used to construct the time-dependent distribution function for an arbitrary initial distribution by convolution. 
The condition (\ref{eq:exdist_coni}) and the orthonormality relation among the wavefunctions determine the coefficient $c_{n}$ and the solution is thus given by 
\begin{equation} 
	f (\phi, \, N) 
	= \frac{ \displaystyle \exp \qty[ - \frac{4 \pi^{2}}{3 H^{4}} V (\phi) ] }{ \displaystyle \exp \qty[ - \frac{4 \pi^{2}}{3 H^{4}} V (\phi_{0}) ] } 
	\sum_{n = 0}^{\infty} \widehat{\Psi}_{n} (\phi) \widehat{\Psi}_{n} (\phi_{0}) e^{- \lambda_{n} (N - N_{0})} 
    = \frac{ \widehat{\Psi}_{0} (\phi) }{ \widehat{\Psi}_{0} (\phi_{0}) } 
	\sum_{n = 0}^{\infty} \widehat{\Psi}_{n} (\phi) \widehat{\Psi}_{n} (\phi_{0}) e^{- \lambda_{n} (N - N_{0})} 
	\,\, . 
	\label{eq:exdist_spexps}
\end{equation}
Equation~(\ref{eq:exdist_spexps}) gives a general formula that enables us to construct exact solutions in the stochasic-inflation side from exact wavefunctions obtained in non-relativistic quantum mechanics. 
When the distribution function is constructed from the quantum harmonic oscillator demonstrated in Section~\ref{subsec:hodemo}, Eq.~(\ref{eq:exdist_spexps}) matches with Eq.~(\ref{eq:pre_dem_distf}) provided the normalised wavefucntions (\ref{eq:pre_dem_nomwf}). 
In particular, the distribution (\ref{eq:exdist_spexps}) approaches its stationary state as $N \to \infty$, 
\begin{equation}
	f_{\infty} (\phi) 
	= \lim_{N \to \infty} f (\phi, \, N) 
	= [ \widehat{\Psi}_{0} (\phi) ]^{2} 
	\,\, , 
\end{equation}
as only the zeromode survives while all the excited modes decay, see also Eq.~(\ref{eq:pre_dist_eqb}). 
The stationary distribution does not depend on the initial field value. 

The potential for the test field in stochastic-inflation side can be obtained by solving the Riccati equation (\ref{eq:pre_spexps}) under some specific condition on the integration constant, as mentioned below Eq.~(\ref{eq:pre_spexps}), but it is more accessible to introduce the following function (sometimes called the super-potential):
\begin{equation}
	W (\phi) 
	\coloneqq - \sqrt{ \frac{H^{2}}{8 \pi^{2}} } \, \dv{\phi} \ln \widehat{\Psi}_{0} (\phi) 
	= - A^{1/4} \dv{z} \ln \widehat{\Psi}_{0} (\phi) 
	\,\, . 
    \label{eq:cat_spp}
\end{equation}
It reads 
\begin{equation}
	\frac{V (\phi)}{H^{4}} 
	= \frac{3}{4 \pi^{2}} \sqrt{ \frac{8 \pi^{2}}{H^{2}} } \int \dd \phi \, W (\phi) 
	= \frac{3}{4 \pi^{2}} \ln \qty[ \frac{ \widehat{\Psi}_{0} (0) }{ \widehat{\Psi}_{0} (\phi) } ]  
	\,\, . 
    \label{eq:cat_spotinf}
\end{equation}
The integration constant is chosen so that $V (\phi)$ vanishes at its origin, $\phi = 0$. 

\subsection{Statistical quantities for deficient oscillator I}

Let us first consider the $\mathscr{D} = \{ 1, \, 2 \}$ quantum deficient oscillator discussed in Section~\ref{subsec:dosc1}. 
The ground-state wavefunction determines the stationary distribution via Eq.~(\ref{eq:df1_wf_nom}) for $n=0$:
\begin{equation}
	f_{\infty} (\phi) 
	= [ \widehat{\Psi}_{0} (\phi) ]^{2} 
	= 2 \qty( \frac{\sqrt{A}}{\pi} \frac{8 \pi^{2}}{H^{2}} )^{1/2} 
    \frac{e^{- z^{2}}}{(2 z^{2} + 1)^{2}} 
	\,\, . 
    \label{eq:df1_stdist}
\end{equation}
In addition, the general time-dependent distribution at arbitrary time $N \geq N_{0}$ is constructed from Eqs.~(\ref{eq:df1_wf_nom}) for $n \ge 1$, and expressed as 
\begin{equation}
	\frac{ f (\phi, \, N) }{ A^{1/4} } 
	= \sqrt{ \frac{8 \pi^{2}}{H^{2}} } \, \frac{e^{- z^{2}}}{(2 z^{2} + 1)^{2}} 
    \qty{ 
        \frac{2}{\sqrt{\pi}} 
        + \sum_{n = 1}^{\infty} 
        \frac{8}{ (n+2) \Gamma (n / 2) \Gamma [ (n+1)/2 ] } 
        \widehat{u}_{n} (z) \widehat{u}_{n} (z_{0}) e^{- 2 (n + 2) \cdot \sqrt{A} \, (N - N_{0})} 
    } 
	\,\, . 
    \label{eq:df1_pdf_insum}
\end{equation}
Note that the ground-state contribution is separated since in the deficient oscillator model the $n = 0$ state and the $n \geq 1$ states have no unified expression in our construction. 
For $z_{0} = 0$, the infinite summation can be reduced to a closed form by virtue of Mehler's summation formula (\ref{eq:pre_ex_mehler}), essentially reproducing the result in Ref.~\cite{1982PhLA...91..396H},\footnote{
    Note, however, typos are found in the reference. 
    Additional (or redundant) $\pi$'s should be multiplied (divided) to correct the expression. 
} 
\begin{align}
	\frac{ f (\phi, \, N) }{ A^{1/4} }  
	&= \sqrt{ \frac{8 \pi^{2}}{H^{2}} } \, \frac{1}{(2 z^{2} + 1)^{2}} 
    \left\{ 
        ( 2 z^{2} + 1 ) \frac{\exp [ - 3 \sqrt{A} (N - N_{0})]}{\sqrt{2 \pi \sinh [ 2 \sqrt{A} \, (N - N_{0}) ]}} 
    \right. 
    \notag \\ 
    &\quad \quad ~~~ \left. 
        + \frac{4}{\sqrt{2 \pi}} \sqrt{ \sinh [ 2 \sqrt{A} \, (N - N_{0}) ] } \exp [ - \sqrt{A} \, (N - N_{0}) ] 
        \vphantom{ \frac{\exp [ - 3 \sqrt{A} (N - N_{0})]}{\sqrt{2 \pi \sinh [ 2 \sqrt{A} \, (N - N_{0}) ]}} }
    \right\} 
    \exp \qty{ 
        - \frac{z^{2}}{1 - \exp [ - 4 \sqrt{A} \, (N - N_{0}) ] } 
    } 
	\,\, . 
    \label{eq:df1_pdf_cl}
\end{align}

\begin{figure}
	\centering
	\includegraphics[width = 0.8\linewidth]{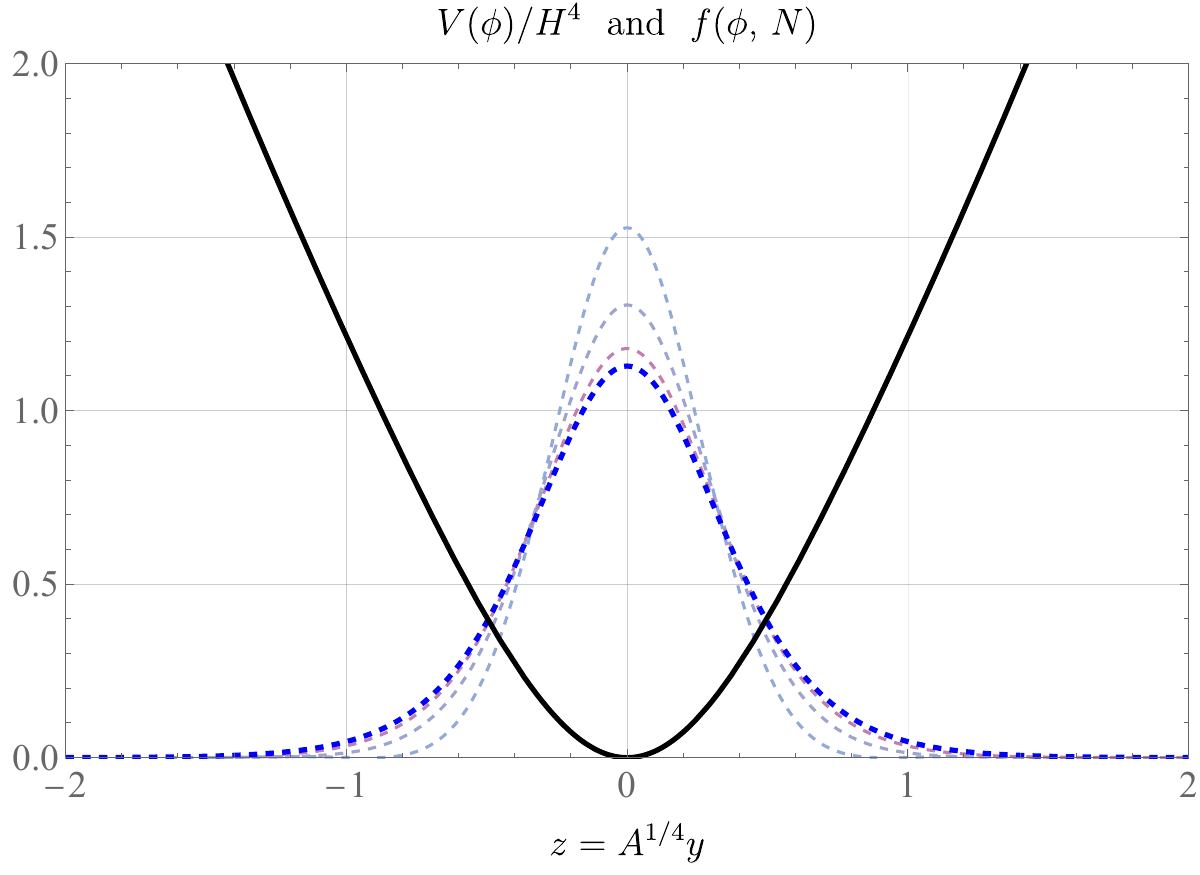}
	\caption{
		The scalar-field potential (\ref{eq:df1_pot_cos}) (solid curve) and the distribution function of the scalar field (dashed curves) for $\sqrt{A} \, (N - N_{0}) = 0.05, \, 0.1, \, 0.2$, and $N \to \infty$ (thick dashed blue curve). 
		The potential is rescaled by ten times for illustrative purpose. 
	}
	\label{fig:df1_pdf}
\end{figure}

The corresponding expression of statistical moments is omitted due to its messiness but can also be obtained analytically in terms of exponentials, hyperbolic functions, and the exponential integrals. 
In late-time limit, it approaches to 
\begin{equation}
	\expval{ z^{n} }_{\infty} 
	\coloneqq \lim_{N \to \infty} \expval{ z^{n} } (N) 
	= \frac{1}{2 \sqrt{\pi}} \qty[ 
        \frac{n \sqrt{e}}{2} E_{(n - 1)/2} \qty( \frac{1}{2} ) - 1 
    ] \Gamma \qty( \frac{n-1}{2} ) 
	\,\, , 
    \qquad 
    n = 0, \, 2, \, \dots 
    \,\, , 
    \label{eq:df2_statmom_ltl}
\end{equation}
where we have introduced the generalised exponential integral  
\begin{equation}
    E_{s} (z) 
    \coloneqq 
    \int_{1}^{\infty} \dd t \, \frac{e^{- z t}}{t^{s}} 
    \,\, . 
\end{equation}
From Eq.~(\ref{eq:df2_statmom_ltl}), the kurtosis of the stationary distribution can be measured, for instance, as 
\begin{equation*}
    \frac{ \expval{ z^{4} }_{\infty} }{ \expval{ z^{2} }_{\infty}^{2} } - 3 
    = \frac{
        2 \sqrt{e} \, E_{3/2} (1/2) - 1 
    }{
        \qty[ \sqrt{e} \, E_{1/2} (1/2) - 1 ]^{2} 
    } - 3 
    \approx 0.89 
    \,\, , 
\end{equation*}
which means that the stationary distribution (\ref{eq:df1_stdist}) is \textit{leptokurtic}, contrary to the \textit{platykurtic} $\lambda \phi^{4}$ theory (in that case the kurtosis is $\expval{ \phi^{4} }_{\infty} / \expval{ \phi^{2} }^{2}_{\infty} - 3 \approx - 0.81$~\cite{Honda:2023unh}). 
On the other hand, the statistical moments vanish for $n = 1, \, 3, \, \dots$ since the model has $\mathbb{Z}_{2}$-symmetry and in the stationary state the initial location of the field is irrelevant. 

The potential in which the scalar field is confined can be read off from Eq.~(\ref{eq:df1_stdist}), or equivalently from Eq.~(\ref{eq:cat_spotinf}), as 
\begin{equation}
	\frac{V (\phi)}{H^{4}} 
	= \frac{3}{4 \pi^{2}} \qty[ 
		\frac{z^{2}}{2} + \ln ( 2 z^{2} + 1 ) 
	] 
	\,\, . 
    \label{eq:df1_pot_cos}
\end{equation}
Around $z = 0$, it behaves as $V (\phi) / H^{4} =  (3 / 4 \pi^{2}) \cdot [5 z^{2} / 2 + \mathcal{O} (z^{4}) ] = (3 / 4 \pi^{2}) [ 5 z^{2} / 2 - 2 z^{4} + \mathcal{O} (z^{6}) ]$. 
Figure~\ref{fig:df1_pdf} shows the relaxation behaviour (\textit{i.e.}~time evolution) of the distribution function (\ref{eq:df1_pdf_cl}) together with the potential (\ref{eq:df1_pot_cos}). 
The expression for the distribution, given by the infinite summation in (\ref{eq:df1_pdf_insum}), matches the curves displayed in the figure when a sufficient number of terms are included, except for small values of $\sqrt{A} \, (N - N_{0})$ where the distribution is concentrated. 
This is because more and more modes must be taken into account at earlier time, while in $N \to \infty$ limit only the zeromode survives. 

\subsection{Statistical quantities for deficient oscillator II}

Next, the exact distribution function for the scalar field corresponding to our second model of the quantum deficient oscillator is constructed. 
In this case, the stationary and time-dependent distributions are given by 
\begin{equation}
    f_{\infty} (\phi) 
    = [ \widehat{ \Psi }_{0} (\phi) ]^{2} 
    = \frac{A^{1/4}}{\sqrt{\pi}} \sqrt{ \frac{8 \pi^{2}}{H^{2}} } \, \frac{24 e^{- z^{2}}}{(4 z^{4} + 3)^{2}} 
    \qty( z^{2} + \frac{1}{2} )^{2} 
    \,\, , 
\end{equation}
and 
\begin{align}
    \frac{ f (\phi, \, N) }{ A^{1/4} } 
    &= \frac{1}{\sqrt{\pi}} \sqrt{ \frac{8 \pi^{2}}{H^{2}} } \, \frac{e^{- z^{2}}}{(4 z^{4} + 3)^{2}} 
    \left\{ 
        24 \qty( z^{2} + \frac{1}{2} )^{2} 
        + 16 \frac{2 z^{2} + 1}{2 z_{0}^{2} + 1} \qty( z^{3} + \frac{3}{2} z ) \qty( z_{0}^{3} + \frac{3}{2} z_{0} ) e^{- 2 \sqrt{A} \, (N - N_{0}) } 
    \right. 
    \notag \\ 
    &\quad \qquad\qquad\qquad\qquad\quad~ \left. 
        + \, \frac{2 z^{2} + 1}{2 z_{0}^{2} + 1} 
        \sum_{n = 2}^{\infty} \frac{2^{n + 4}}{(n + 1) (n + 2) \Gamma (n - 1)} 
        \widehat{u}_{n} (z) \widehat{u}_{n} (z_{0}) 
        e^{- 2 (n + 2) \cdot \sqrt{A} \, (N - N_{0})} 
    \right\} 
    \,\, , 
    \label{eq:df2_pdf_insum}
\end{align}
respectively.
From Eq.~(\ref{eq:df2_pdf_insum}), the time evolution of the statistical moments can also be calculated. 
In late-time limit, it goes to\footnote{
    For each $n$, the statistical moment (\ref{eq:df2_statm_eqb}) can also be expressed in terms of the Fresnel integrals~\cite{NIST:DLMF} 
    \begin{equation*}
        C (z) 
        \coloneqq \int_{0}^{z} \dd t \, \cos \qty( \frac{\pi}{2} t^{2} ) 
        = z \cdot {}_{1} F_{2} \qty( 
            \begin{matrix} 1/4 \\ 5/4, \, 1/2 \end{matrix} ~ \middle| ~ - \frac{\pi^{2}}{16} z^{4} 
        ) 
        \,\, , 
        \qquad 
        S (z) 
        \coloneqq \int_{0}^{z} \dd t \, \sin \qty( \frac{\pi}{2} t^{2} ) 
        = \frac{\pi}{6} z^{3} \cdot {}_{1} F_{2} \qty( 
            \begin{matrix} 3/4 \\ 7/4, \, 3/2 \end{matrix} ~ \middle| ~ - \frac{\pi^{2}}{16} z^{4} 
        ) 
        \,\, . 
    \end{equation*}
} 
\begin{align}
    \expval{ z^{n} }_{\infty} 
    &= \frac{\sqrt{\pi / 2}}{\sin [ (n + 1) \pi / 4 ] \cos [ (n + 1) \pi / 4 ] } 
    \left\{ 
        \frac{2^{n/2}}{128} \cdot 3 \sqrt{\pi} \left[ 
            16 \cdot {}_{1} \widetilde{F}_{2} \qty( ~ 
                \begin{matrix} 2 \\ (5 - n)/4, \, (7-n)/4 \end{matrix} \,~ \middle| ~ - \frac{3}{16} 
            ) 
        \right. 
    \right. 
    \notag \\ 
    &\qquad \left. 
        - 8 \cdot {}_{1} \widetilde{F}_{2} \qty( 
            \begin{matrix} 2 \\ (7-n)/4, \, (9-n)/4 \end{matrix} ~ \middle| ~ - \frac{3}{16} 
        ) 
        + {}_{1} \widetilde{F}_{2} \qty( 
            \begin{matrix} 2 \\ (9-n)/4, \, (11-n)/4 \end{matrix} ~ \middle| ~ - \frac{3}{16} 
        ) 
    \right] 
    \notag \\ 
    &\quad 
    \left. 
        + \frac{n}{2^{n/2}} \cos \qty( \frac{\sqrt{3}}{2} + \frac{3 n + 7}{12} \pi ) 
    \right\} 
    \,\, , 
    \qquad 
    n = 0, \, 2, \, \dots 
    \,\, . 
    \label{eq:df2_statm_eqb}
\end{align}
In Eq.~(\ref{eq:df2_statm_eqb}), the regularised hypergeometric function has been introduced by 
\begin{equation}
    {}_{p} \widetilde{F}_{q} \qty( 
        \begin{matrix} a_{1}, \, \dots, \, a_{p} \\ b_{1}, \, \dots, \, b_{q} \end{matrix} ~ \middle| ~ z 
    ) 
    \coloneqq 
    \frac{
        \displaystyle {}_{p} F_{q} \qty( 
        \begin{matrix} a_{1}, \, \dots, \, a_{p} \\ b_{1}, \, \dots, \, b_{q} \end{matrix} ~ \middle| ~ z 
    ) 
    }{
        \Gamma (b_{1}) \cdot \cdots \cdot \Gamma (b_{q}) 
    } 
    \,\, . 
\end{equation}

The potential for the scalar field reads 
\begin{equation}
	\frac{V (\phi)}{H^{4}} 
	= \frac{3}{4 \pi^{2}} \qty[ 
		\frac{z^{2}}{2} + \ln \qty( \frac{4 z^{4} + 3}{6 z^{2} + 3} ) 
	] 
	\,\, . 
    \label{eq:df2_pot_cos}
\end{equation}
This is a double-well potential, and therefore, the explicit expression of the distribution function (\ref{eq:df2_statm_eqb}) enables one to analytically analyse the scalar-field dynamics in the double-well potential, tracking the time evolution from the beginning to the end including the transient states, while keeping the parametric dependence transparent. 
The Fokker--Planck equation with a piecewise double-well potential
was studied in \textit{e.g.}~Refs.~\cite{PhysRevA.48.4062, CALDAS201492}, where the quantum-mechanical system is solved in terms of the parabolic cylinder functions. 

\begin{figure}
	\centering
	\includegraphics[width = 0.8\linewidth]{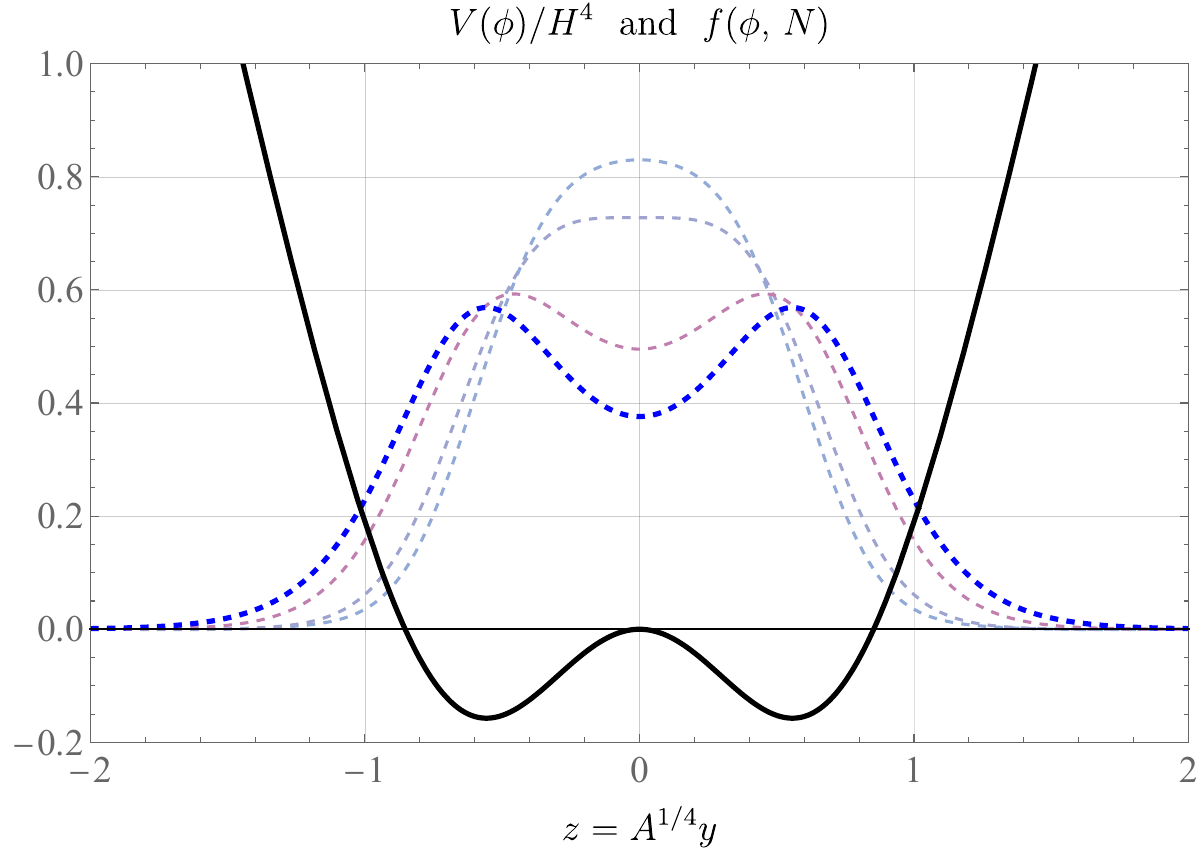}
	\caption{
		The scalar-field potential (\ref{eq:df2_pot_cos}) (solid curve) and the distribution function of the scalar field (dashed curves) for $\sqrt{A} \, (N - N_{0}) = 0.08, \, 0.1, \, 0.2$, and $N \to \infty$ (thick dashed blue curve). 
		The potential is rescaled by ten times for illustrative purpose. 
	}
	\label{fig:df2_pdf}
\end{figure}

Figure~\ref{fig:df2_pdf} shows the time evolution of the distribution function, Eq.~(\ref{eq:df2_pdf_insum}), truncated at $n_{\rm max} = 20$. 
Since at earlier times $n_{\rm max} = \mathcal{O} (10)$ is insufficient to closely reproduce the exact behaviour of the distribution function, it is displayed for relatively later times where our choice of the truncation order is sufficient to reproduce the reasonable behaviour. 
The potential is also shown in the figure, and the single-peaked concentrated distribution at earlier times gradually relaxes in a way that, as time passes, it splits into the two peaks accordingly. 

Before moving to the next and last subsection, let us comment on the identification of $A$ with the mass and on the typical timescales. 
Going back to the quantum harmonic oscillator potential (\ref{eq:pre_ex_infv}), it can be rewritten as $V (\phi) = (6 \sqrt{A} \, H^2 / 2) \phi^{2}$, which enables us to relate $A$ with the mass of the scalar field by $(m / H)^{2} = 6 \sqrt{A}$. 
Under this identification, the timescale in which the mean of the field relaxes is given by 
\begin{equation}
    N_{1}  
    \coloneqq \frac{1}{2 \sqrt{A}} 
    = \frac{3 H^{2}}{m^{2}} 
    \,\, , 
    \label{eq:rxtime_n1}
\end{equation}
as is known~\cite{Enqvist:2012xn, Hardwick:2017fjo, Tokeshi:2024kuv}, implying that a sufficient long time is needed to settle down to the stationary state provided that the mass of the scalar field is light compared to the Hubble scale. 
The timescale for the variance is given by $N_{2} \coloneqq 1 / 4 \sqrt{A} = 3 H^{2} / 2 m^{2}$. 
A similar discussion can be made for the transformed anharmonic oscillators discussed in Section~\ref{sec:esnv}, but it is not straightforward to obtain the analytical formula such as Eq.~(\ref{eq:rxtime_n1}). 
Numerical calculations are instead required to find out the typical timescales of the statistical quantities settling on the stationary values. 

\subsection{Different choices of energy levels extracted}
\label{subsec:gallery}

\begin{figure}
    \centering
    \includegraphics[width = 0.995\linewidth]{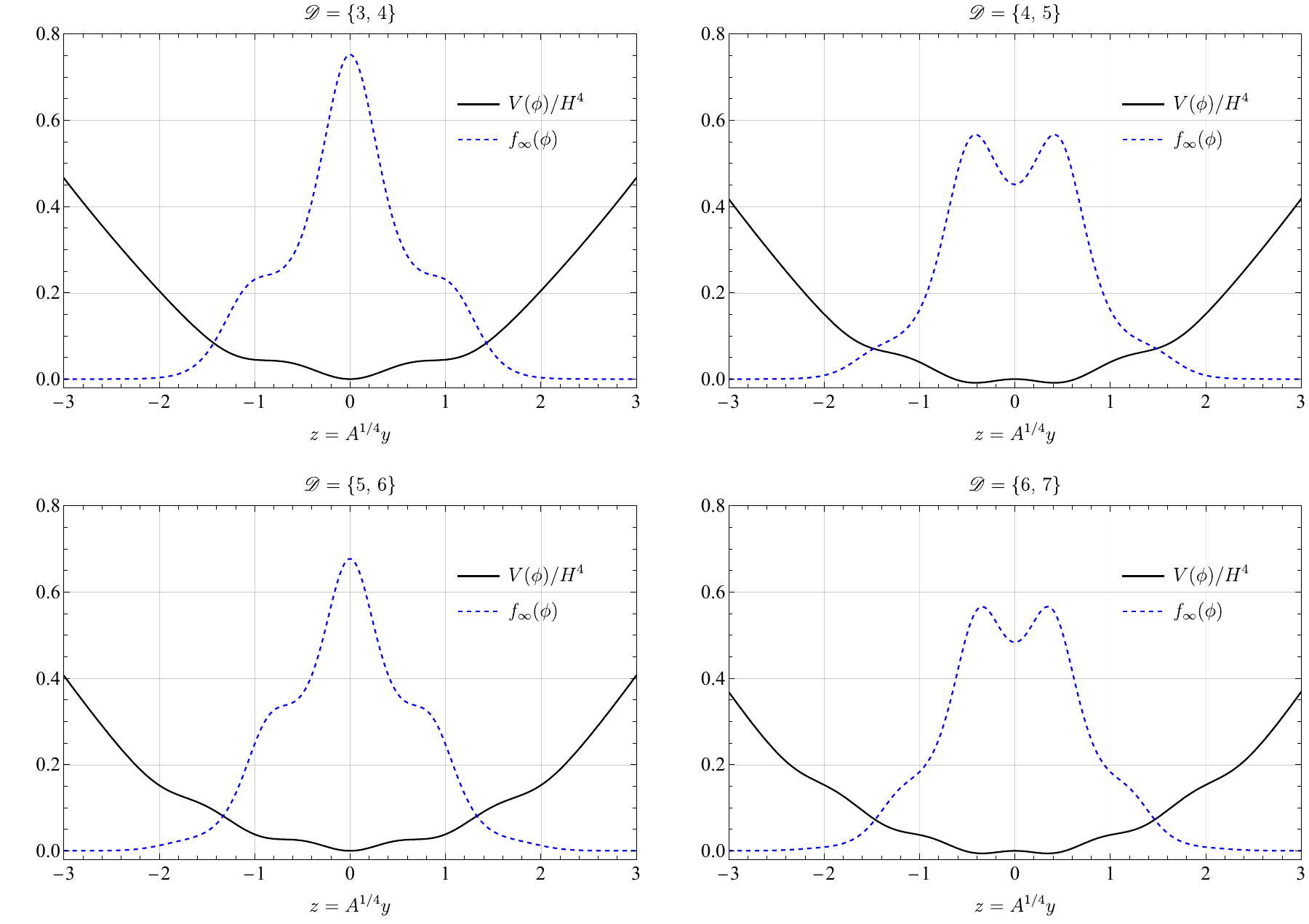}
    \caption{
        Several exactly solvable potentials (solid black curve) of the scalar field $\phi$ in the Fokker--Planck side, together with the stationary distribution (dashed blue curve). 
        In all the figures, the stationary distributions are rescaled in such a way that $(\sqrt{ H^{2} / 8 \pi^{2}} / A^{1/4} ) f_{\infty} (\phi)$ is actually plotted. 
        The potential has its single bottom at the origin for $d_{1} = 3$ and $d_{1} = 5$, while for $d_{1} = 4$ and $d_{1} = 6$ the double-well nature can be observed. 
    }
    \label{fig:dfgen_fig}
\end{figure}

The general procedure of the Krein--Adler transformation summarised in Section~\ref{subsec:katrf} enables us to construct infinitely many exactly solvable potentials in stochastic inflation. 
Though the closed-form statistical quantities such as the distribution and correlation functions can hardly be expected in most cases, analytical expressions leaving an infinite summation can still be used for numerical implementation, except for small $N - N_{0}$, thanks to the quickly converging contributions from the higher modes with the exponentially suppressing temporal factor. 
Restricting ourselves to the case where $\mathscr{D} = \{ d_{1}, \, d_{2} = d_{1} + 1 \}$, \textit{i.e.}~the newly constructed exactly solvable models that can be obtained by deleting a pair of the neighbouring two states in the quantum harmonic oscillator, several corresponding models in the de Sitter side are displayed in this section. 
Construction of the time-evolving distribution function and generalisation for cases where $\ell \neq 2$ are both straightforward as will be seen soon, following the procedure summarised and demonstrated in Section~\ref{sec:esnv}. 

Let us here give several exactly solvable potentials in the Fokker--Planck side. 
For $3 \leq d_{1} \leq 6$, the potentials constructed by Eqs.~(\ref{eq:cat_kwfnom}) and (\ref{eq:cat_spotinf}) read 
\begin{itemize}
	\item $\mathscr{D} = \{ 3, \, 4 \}$ 
		\begin{equation}
			\frac{V (\phi)}{H^{4}} 
			= \frac{3}{4 \pi^{2}} \qty[ \frac{z^{2}}{2} + \ln \qty( 
			\frac{8 z^{6} - 12 z^{4} + 18 z^{2} + 9}{ 12 z^{4} + 9 } 
		) ] 
        \,\, . 
		\end{equation}
	
	\item $\mathscr{D} = \{ 4, \, 5 \}$ 
		\begin{equation}
			\frac{V (\phi)}{H^{4}} 
			= \frac{3}{4 \pi^{2}} \qty[ \frac{z^{2}}{2} + \ln \qty( 
			\frac{16 z^{8} - 64 z^{6} + 120 z^{4} + 45}{ 40 z^{6} - 60 z^{4} + 90 z^{2} + 45 } 
		) ] 
        \,\, . 
		\end{equation}

    \item $\mathscr{D} = \{ 5, \, 6 \}$ 
		\begin{equation}
			\frac{V (\phi)}{H^{4}} 
			= \frac{3}{4 \pi^{2}} \qty[ \frac{z^{2}}{2} + \ln \qty( 
			\frac{ 32 z^{10} - 240 z^{8} + 720 z^{6} - 600 z^{4} + 450 z^{2} + 225}{ 80 z^{8} - 320 z^{6} + 600 z^{4} + 225 } 
		) ] 
        \,\, . 
		\end{equation}

    \item $\mathscr{D} = \{ 6, \, 7 \}$ 
		\begin{equation}
			\frac{V (\phi)}{H^{4}} 
			= \frac{3}{4 \pi^{2}} \qty[ \frac{z^{2}}{2} + \ln \qty( 
			\frac{ 64 z^{12} - 768 z^{10} + 3600 z^{8} - 6720 z^{6} + 6300 z^{4} + 1575 }{ 224 z^{10} - 1680 z^{8} + 5040 z^{6} - 4200 z^{4} + 3150 z^{2} + 1575 } 
		) ] 
        \,\, . 
		\end{equation}
\end{itemize}
Figure~\ref{fig:dfgen_fig} shows those potentials together with the stationary distributions, \textit{i.e.}~the corresponding wavefunctions squared. 
It can be seen that the larger $d_{1}$ becomes, the more wavy behaviour the potential shows. 
Also, the potential $V (\phi)$ for the scalar field $\phi$ is always symmetric and has a single global minimum for $d_{1} = 3$ or $d_{1} = 5$, and the stationary distribution is peaked at $z = 0$, small bumps being also present at the both sides of the origin. 
On the other hand, the potential has two minima for $d_{1} = 4$ or $d_{1} = 6$, realising the double wells, and the stationary distribution is endowed with its double peak accordingly. 
The scalar-field potentials and the corresponding distribution functions for other choices of $d_{1}$ can also be obtained in the same way. 
Generalising the above discussion, for an odd $d_{1}$ in $\mathscr{D} = \qty{ d_{1}, \, d_{1} + 1 }$ the global minimum of the inflationary potential is single, while for an even $d$ it is essentially double-well, which allows analyses of various phenomenologies to be studied analytically. 

\begin{figure}
    \centering
    \includegraphics[width=0.995\linewidth]{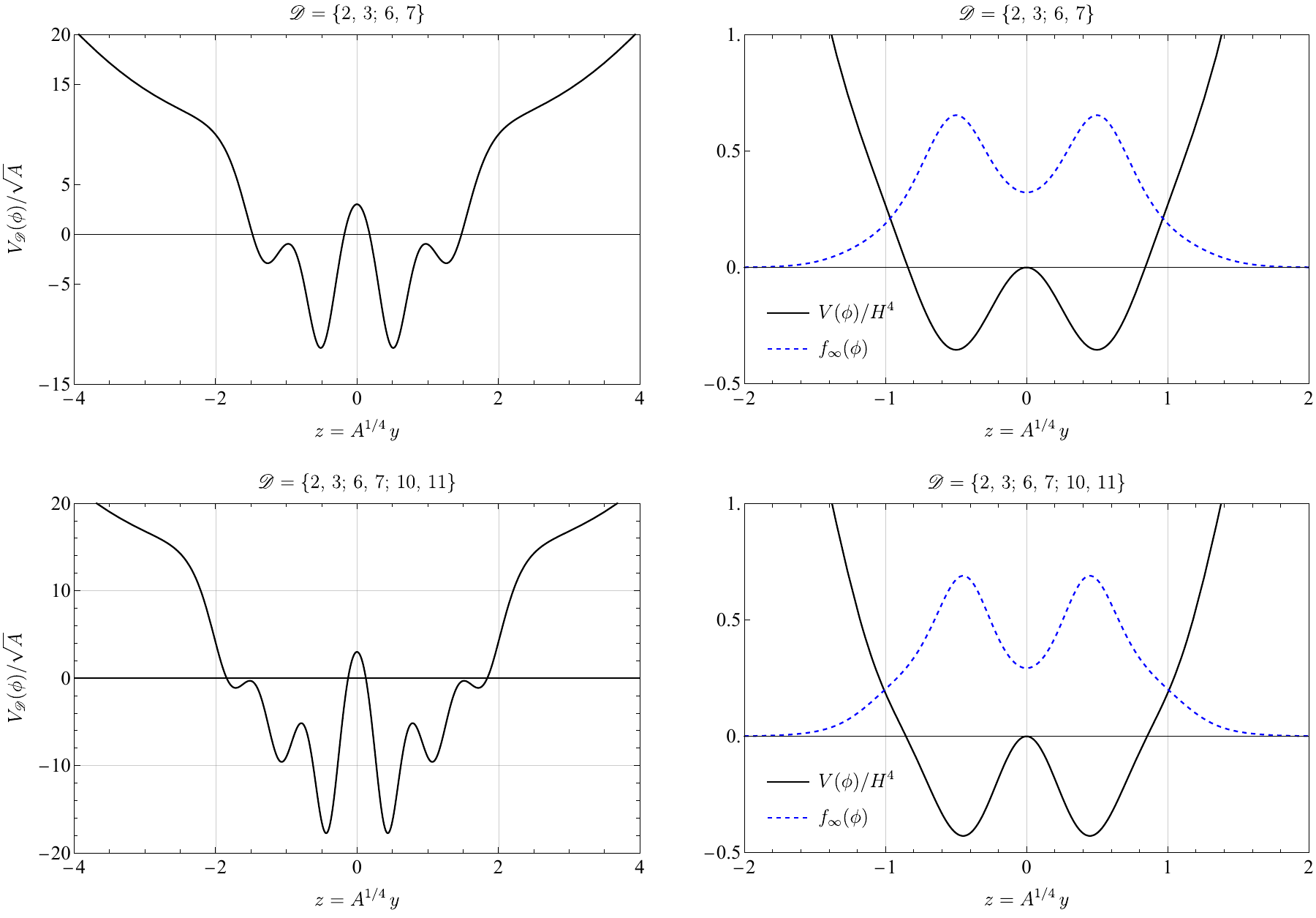}
    \caption{
        The Schr\"{o}dinger potentials (\textit{left}) and the corresponding Fokker--Planck potentials together with the stationary distributions (\textit{right}). 
        The top panels are for $\mathscr{D} = \qty{ 2, \, 3 ; \, 6, \, 7 }$ and the bottom for $\mathscr{D} = \qty{ 2, \, 3 ; \, 6, \, 7; \, 10, \, 11 }$. 
        In the right panels, both the potentials and the stationary distributions are rescaled in such a way that $(4 \pi^{2} / 3) V (\phi) / H^{4}$ and $(\sqrt{ H^{2} / 8 \pi^{2}} / A^{1/4} ) f_{\infty} (\phi)$ are actually plotted. 
    }
    \label{fig:rac}
\end{figure}

Other exactly solvable double-well potentials in the stochastic-inflation side can also be constructed from the transformed quantum deficient oscillators with $\ell > 2$. 
The left and right panels of Figure~\ref{fig:rac} respectively show the Schr\"{o}dinger potential (constructed by the general formula, Eq.~(\ref{eq:cat_def})) and the corresponding Fokker--Planck potential (constructed from Eq.~(\ref{eq:cat_spotinf}) through the ground-state wavefunction obtained by Eq.~(\ref{eq:katrf_wfgen})) together with the stationary distributions. 
The difference between the upper and lower panels of the same figure is the choice of $\mathscr{D}$, 
\begin{align}
    \ell = 4 
    & &\text{and} 
    & &\mathscr{D} &= \qty{ 2, \, 3; \, 6, \, 7 }, 
    & &\textit{i.e.}\quad\qty( d_{1} = 2, \, d_{2}; \, d_{3} = 6, \, d_{4} )
    \,\, , 
    \notag \\ 
    \ell = 6 
    & &\text{and}
    & &\mathscr{D} &= \qty{ 2, \, 3; \, 6, \, 7; \, 10, \, 11 }, 
    & &\textit{i.e.}\quad\qty( d_{1} = 2, \, d_{2}; \, d_{3} = 6, \, d_{4}; \, d_{5} = 10, \, d_{6} ) 
    \,\, , 
    \notag 
\end{align}
respectively, where $d_{2} = d_{1} + 1$, $d_{4} = d_{3} + 1$, and $d_{6} = d_{5} + 1$. 
Those indicate the extracted (deficient) energy levels in the quantum harmonic oscillator. 
The stationary distribution (blue dashed curve) is properly normalised. 
Though many exactly solvable double-well models can be constructed in the stochastic-inflation side, a squishy characteristic of the Schr\"{o}dinger potential is not necessarily handed down to the behaviour in the Fokker--Planck potential. 
The latter behaves more modestly in most cases, as can be seen in Figure~\ref{fig:rac}. 
On the other hand, the $\mathbb{Z}_{2}$-symmetric nature of the potentials can always be observed both in the transformed potentials in quantum mechanics and the corresponding diffusive potentials. 
Comparing the two cases with $\mathscr{D} = \qty{ 2, \, 3; \, 6, \, 7 }$ and $\mathscr{D} = \qty{ 2, \, 3; \, 6, \, 7; \, 10, \, 11 }$ (the two panels in the left or right side in Figure~\ref{fig:rac}), it can be noticed that the additional extraction of the energy levels ($n = 10$ and $11$) does not affect the behaviours around $z = 0$ significantly. 
This is because the lower excited states such as $n = 2$ and $n = 3$ are mainly related to nearly the bottom of the potential, while the higher excited states to the steeper part of the potential, as can be seen also in Figure~\ref{fig:dfgen_fig}. 
In summary, Figure~\ref{fig:rac} demonstrates that, though many quantitatively different models can be constructed by the Krein--Adler transformations, the \textit{additional} extraction of higher excited states in the quantum harmonic oscillator would not necessarily generate a new solvable system that is qualitatively different. 
Nevertheless, qualitatively different models can still arise in certain cases, as we discuss below. 

\begin{figure}
    \centering
    \includegraphics[width=0.995\linewidth]{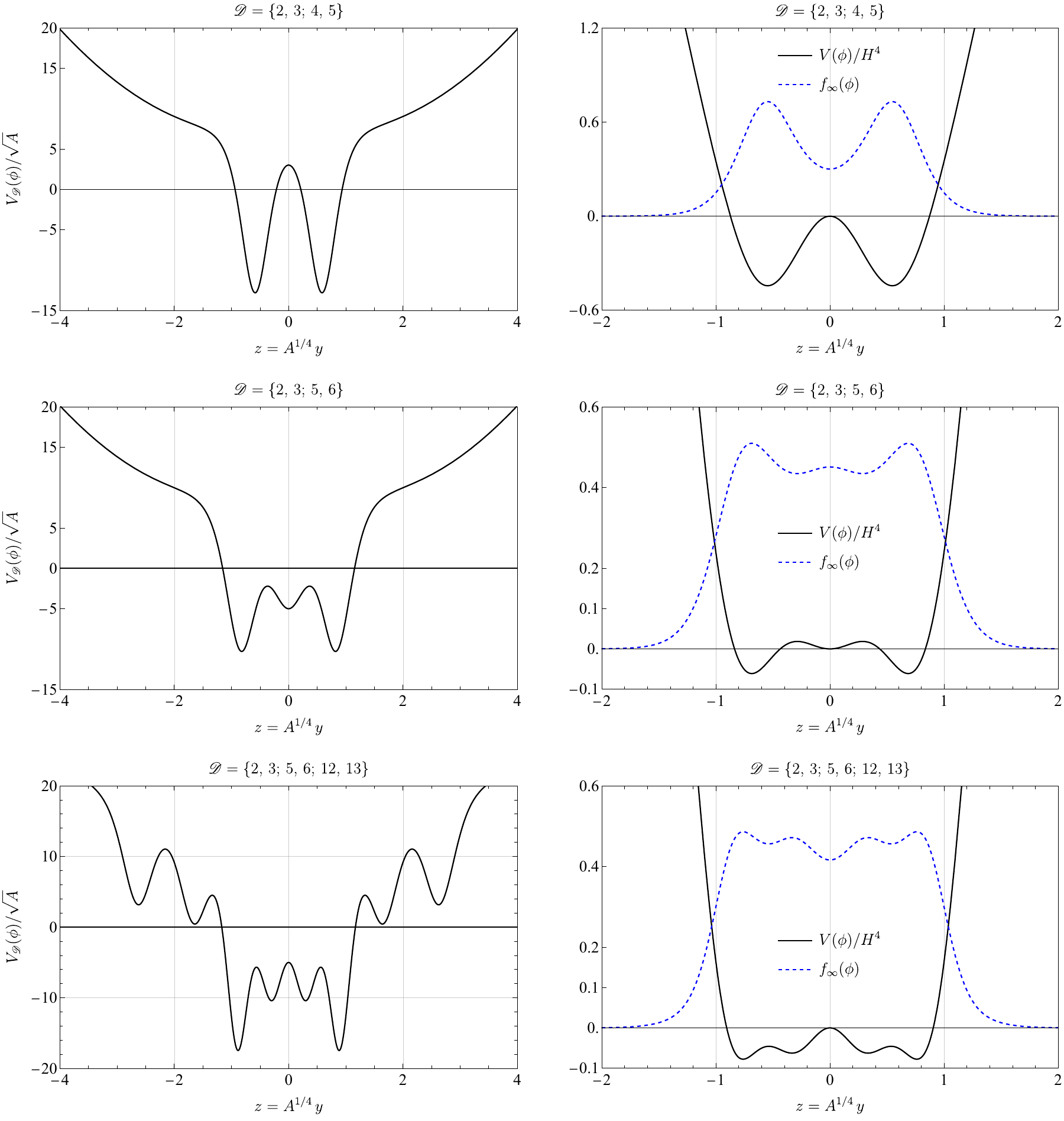}
    \caption{
        The Schr\"{o}dinger potentials (\textit{left}) and the corresponding Fokker--Planck potentials together with the stationary distributions (\textit{right}). 
        The top panels are for $\mathscr{D} = \qty{ 2, \, 3 ; \, 4, \, 5 }$ and the middle for $\mathscr{D} = \qty{ 2, \, 3 ; \, 5, \, 6 }$, while the bottom is for $\mathscr{D} = \qty{ 2, \, 3 ; \, 5, \, 6; \, 12, \, 13 }$. 
        In the right panels, as was done in Figure~\ref{fig:rac}, all the potentials and the stationary distributions are rescaled in such a way that $(4 \pi^{2} / 3) V (\phi) / H^{4}$ and $(\sqrt{ H^{2} / 8 \pi^{2}} / A^{1/4} ) f_{\infty} (\phi)$ are actually plotted. 
    }
    \label{fig:rac2}
\end{figure}

Last but not least, exactly solvable multiple-well potentials can be constructed as well, both in quantum mechanics and stochastic inflation by employing the Krein--Adler transformation again with $\ell > 2$. 
Figure~\ref{fig:rac2} demonstrates that, in the stochastic-inflation side, a potential endowed with an arbitrary number of wells can be constructed. 
The Schr\"{o}dinger and Fokker--Planck potentials are displayed in the left and right panels, respectively. 
The difference among the top, middle, and bottom panels is the choice of $\mathscr{D}$. 
The potential of the quantum deficient oscillator in the top left panel can be obtained by deleting the subsequent four energy levels in the quantum harmonic oscillator, to the right of which the double-well corresponding Fokker--Planck potential is displayed. 
From the point of view of the stochastic-inflation side, the qualitative outcome is expected to be shared by the models displayed in Figure~\ref{fig:rac} since all the potentials have their double-well nature, while quantitative differences may be anticipated. 
On the other hand, the middle and bottom panels show that not only a class of exactly solvable models with the double wells (during inflation) can be constructed by the Krein--Adler transformation, but also models with more than three wells can be constructed, to which exact statistical quantities such as the distribution and correlation functions are still available. 
For instance, one single local minimum and the two global minima can be observed in the middle right panel of Figure~\ref{fig:rac2}, 
which thus allows us to analytically analyse \textit{e.g.}~vacuum decay in de Sitter spacetime. 
The potential of this triple-well model is in particular given by 
\begin{equation}
    \frac{V (\phi)}{H^{4}} 
    = \frac{3}{4 \pi^{2}} \qty[ 
        \frac{z^{2}}{2} 
        + \ln \qty( 
            \frac{
                32 z^{10} + 80 z^{8} + 160 z^{6} + 150 z^{2} + 75 
            }{
                120 z^{6} + 300 z^{4} + 150 z^{2} + 75 
            }
        ) 
    ] \,\, . 
\end{equation}
Those figures demonstrate that the extraction of the two pairs of the energy levels in the quantum harmonic oscillator can give rise to the Fokker--Planck potentials with multiple wells. 
If one considers the quantum-mechanical system in which the three pairs of the energy levels are deleted compared to the quantum harmonic oscillator, the corresponding potential in the stochastic-inflation side with the quadruple wells can be constructed, an example of which is illustrated in the bottom. 
This observation may be generalised to realise a model with an arbitrary number of wells describing the scalar field dynamics in an expanding universe. 

\section{Conclusion}
\label{sec:disc}

Construction of exactly solvable models in theoretical physics can never be underestimated. 
The power-law model~\cite{Lucchin:1984yf} is presumably the most famous exact solution in inflationary cosmology, in which the inflaton field rolls on the exponential potential and the Klein--Gordon field equation can be solved exactly without the slow-roll prescription, keeping the dependence on the field of the Hubble parameter through the Friedmann equation. 
While the background equation of motion in the slow-roll regime can be solved analytically in most cases, it ceases to be true when the stochastic effects are taken into account even with the slow-roll approximation. 
A few exactly solvable models include the ones with the quartic potential~\cite{Yi:1991ub}, the cosmic time $t$ being used instead of the number of $e$-folds $N$~\cite{Finelli:2008zg, Vennin:2015hra}, and the flat-well potential~\cite{Pattison:2017mbe, Ezquiaga:2019ftu, Ando:2020fjm, Pattison:2021oen, Animali:2024jiz, Animali:2025pyf} that realises an ultra-slow-roll phase to be associated possibly with formation of primordial black holes. 

For a non-inflaton light field, on the other hand, the Fokker--Planck equation considerably simplifies due to the ignorance of the variation of the Hubble parameter. 
This allows us to construct a number of exact solutions therein, and a class of them respecting the shape invariance of the corresponding quantum-mechanical potentials is systematically studied in Ref.~\cite{Honda:2024evc}. 
As a natural but important extension of Ref.~\cite{Honda:2024evc}, the present article has examined a variety of exact solutions in stochastic inflation by exploiting the Krein--Adler transformation and thus going beyond the shape-invariant models. 

The key starting point is the fact that the Fokker--Planck equation in stochastic inflation can be mapped to the Schr\"{o}dinger equation in non-relativistic quantum mechanics, and the inverse procedure can also be done. 
This is why Section~\ref{sec:esnv} starts with the potentials, the wavefunctions in quantum-mechanical side including the proper normalisations are constructed by the elementary but explicit method, which can rarely be found in the literature. 
The recurrence relations satisfied by the coefficients in the series-solution ansatz (\ref{eq:df1_ser_fb}) are non-trivially and exactly solved, which enables one to reduce the infinite summations to the closed forms, again non-trivially. 
Each model can be obtained by deleting a set of energy levels in the quantum harmonic oscillator, and can be understood systematically by the theory of the Krein--Adler transformation as summarised in Section~\ref{subsec:katrf}. 
Indeed, different choices of the energy levels extracted in the harmonic oscillator give different exactly solvable models, where the remaining energy eigenvalues are unchanged while the corresponding wavefunctions are modified. 
Though the simplest two models are examined in this article, the Krein--Adler transformation offers a general procedure by which infinitely many exactly solvable models are constructed. 
Those models are said to be ``non-classical'' exactly solvable ones in the sense that they no longer respect the shape invariance and therefore cannot be solved simply by the classical orthogonal polynomials. 

Going back to the stochastic-inflation side in Section~\ref{sec:statds}, exactly solvable models that describe the scalar-field dynamics in de Sitter spacetime are constructed from the normalised wavefunctions that solve the quantum-mechanical models considered in Section~\ref{sec:esnv}. 
The distribution functions are explicitly given, which are exact and describe all the time-dependent behaviour from the initial concentrated one and the transient time-evolving one, to the stationary distribution that can only be achieved non-perturbatively by virtue of the stochastic formalism.  
The time-dependent correlation functions (statistical moments) can be derived from the distribution, but this article restricted itself to the stationary-state calculations for simplicity. 
In the first model (corresponding to the one considered in Section~\ref{subsec:dosc1}), the single-well nature is common in both quantum-mechanical and stochastic-inflation sides. 
On the other hand, the double-well nature of the Krein--Adler-transformed potential (considered in Section~\ref{subsec:dosc2}) is passed down to the potential for the scalar field, providing an exactly solvable double-well model in the de Sitter (inflationary) context. 
It thus opens a window for analytical studies on various subjects including vacuum decay, curvaton phenomenology, and physics of primordial black holes. 
Those demonstrative cases are further extended to other exactly solvable models in Section~\ref{subsec:gallery}, being achieved by considering various choices of the energy levels in the quantum harmonic oscillator that are deleted in the corresponding transformed models. 

In addition to applications of the exact distribution functions and statistical quantities presented in this article to various phenomenologies mentioned above, another interesting future direction would be to study other classes of the exact solutions in quantum mechanics, sometimes realising a non-symmetric scalar-field potential. 
Such cases can be obtained by considering the Krein--Adler transformations to the other exactly solvable models solved in terms of the associated Laguerre or Jacobi polynomials.
Other possible directions include exactly solvable quantum-mechanical potentials solved via exceptional orthogonal polynomials and other multi-indexed orthogonal polynomials~\cite{quesne2008exceptional, ODAKE2009414, ODAKE2010173, sasaki2010exceptional, ODAKE2011164}. 

\acknowledgements

This work was supported by the JSPS KAKENHI Grant Number JP24K22877. 

\bibliographystyle{JHEP}

\providecommand{\href}[2]{#2}\begingroup\raggedright\endgroup

\end{document}